\documentclass[galaxies,article,accept,pdftex,moreauthors]{mdpi}

\usepackage{graphicx} 

\newcommand{\teff}{$T_{\rm eff}$}
\newcommand{\eexc}{$E_{\rm exc}$}
\def\vt{$\xi_{\rm t}$}
\def\kms{$\rm km~s^{-1}$}
\def\ione{\,{\sc i}}
\def\ii{\,{\sc ii}}

\newcommand{\eps}{\log\varepsilon}
\def\fodd{F$_{\rm odd}$}

\newcommand{\SV}{\textsc{SynthVb}}

\newcommand{\vald}{\textsc{VALD}}

\def\sj{SDSS~J1349-0229}
\def\cs{BPS~CS~29512-073} 
\def\sjten{SDSS~J1036+1212}

\firstpage{1} 
\pubvolume{14}
\issuenum{3}
\articlenumber{42}
\pubyear{2026}
\copyrightyear{2026}
\externaleditor{Firstname Lastname}
\datereceived{26 February 2026} 
\daterevised{27 April 2026} 
\dateaccepted{29 April 2026} 
\datepublished{6 May 2026} 
\doinum{}

\Title{Ba Isotope Ratio in CEMP-s and CEMP-rs Stars as a Signature of s-Process and i-Process}

\Author{Tatyana 
 Sitnova *
$^{,\dagger}$ and Lyudmila Mashonkina 
 }

\AuthorNames{Tatyana Sitnova and Lyudmila Mashonkina}

\address[1]{%
Institute of Astronomy, Russian Academy of Sciences
; lima@inasan.ru 
}

\corres{\hangafter=1 \hangindent=1.05em \hspace{-0.82em}Correspondence: sitamih@gmail.com}

\firstnote{\hangafter=1 \hangindent=1.05em \hspace{-0.82em} Current address: Pyatnitskaya St.
, 48, 119017 Moscow, Russia.}  

\abstract{
We present a spectroscopic analysis of three carbon-enhanced metal-poor (CEMP) stars of type CEMP-s and CEMP-rs and determine their non-local thermodynamic equilibrium (NLTE) abundances of Ba and the fractions of the odd Ba isotopes (\fodd). We found \fodd\ = 0.65$_{-0.34}^{+0.35}$ in SDSS~J1349-0229, which is known in the literature as a CEMP-rs star, while the other two stars, BPS~CS~29512-073 and SDSS~J1036+1212,  exhibit lower  \fodd\ = 0.23$_{-0.10}^{+0.19}$ and 0.23$_{-0.11}^{+0.22}$, respectively, and they are known in the literature as CEMP-s stars. The present result supports our earlier finding about distinct \fodd\ in CEMP-s and CEMP-rs stars. For obtaining observational constraints on i-process nucleosynthesis, further NLTE abundance determinations for many chemical elements are required. We provide a tool for generating the lists of Ba\ii\ lines for a given \fodd\ and it is available on GitHub \url{https://github.com/sitamih/ba\_linelist}.
}

\keyword{CEMP-s stars; CEMP-rs stars; s-process; i-process; r-process; high-resolution spectroscopy; abundances; isotope ratios} 

\begin{document}


\section{Introduction}

Neutron (n) capture processes may occur at different conditions and, depending on the neutron flux, are classified as slow (s), rapid (r), and intermediate (i) processes \citep{1957RvMP...29..547B,1977ApJ...212..149C}. In this study, we focus on obtaining observational constraints for s- and i-process nucleosynthesis. The main s-process  operates in low- and intermediate-mass stars  during the asymptotic giant branch (AGB) evolutionary stage \citep{1998ApJ...497..388G,2005ARA&A..43..435H}. For the i-process, specific conditions, such as proton ingestions into a He-burning zone, need to occur for producing higher neutron densities than those typical of the s-process. The supposed i-process sites are metal-poor (MP) AGB stars \citep{2011ApJ...727...89H,2016ApJ...833..181C,2022A&A...667A.155C,2024A&A...684A.206C} and accreting white dwarfs \citep{2019MNRAS.488.4258D}. 
The main observational signature of the s-process is an enhanced Ba abundance, while the i-process is predicted to produce a significant overabundance of Eu in addition to Ba.

Ba is considered an s-element, as 88 \%\ of the solar Ba originates from the s-process, while Eu is referred to as an r-element, with 95 \%\ of the solar Eu originating from the r-process \citep{2020MNRAS.491.1832P}. These properties led to the classification of carbon-enhanced metal-poor (CEMP) stars with enhanced abundances of n-capture elements as CEMP-s and CEMP-rs stars, which trace the s- and i-process, respectively. These stars are members of binary systems \citep{2001AJ....122.1545P,2005ApJ...625..825L,2014MNRAS.441.1217S} and they were enriched by material transferred from their companion stars, which have since evolved into white dwarfs. The chemical composition of the CEMP-s,rs stars is important for constraining the s- and i-process nucleosynthesis. While the s-process is well studied and used in galactic chemical evolution modeling, the i-process is less well understood.

To study the i-process, it is first necessary to identify stars that trace i-process nucleosynthesis. Such stars are expected to show radial velocity variations, as well as enhanced C, Ba, and Eu abundances. Several criteria have been proposed to distinguish CEMP-rs stars from CEMP-s stars, and some of them are listed in Table~\ref{classification}. These criteria are largely formal, and in practice stars rarely fall clearly into distinct categories. 

\begin{table*}
\caption{Classification schemes available in the literature for distinguishing between CEMP-s and CEMP-rs stars.}        
\label{classification} 
\footnotesize        
\centering          
\begin{tabular}{lll}   
\hline      
s & rs &  ref.  \\
 \hline                                                           
{[Ba/Eu]} $>$ 0.5                               &  0.0 $<$ {[Ba/Eu]} $<$ 0.5                     & {\citep{2005ARA&A..43..531B}} \\
{[Ba/Fe]} $>$ 1.0, {[Eu/Fe]} $<$ 1, {[Ba/Eu]} $>$ 0 & {[Ba/Fe]} $>$ 1.0, {[Eu/Fe]} $>$ 1, {[Ba/Eu]} $>$ 0 & {\citep{2006A&A...451..651J}} \\
{[Ba/Fe]} $>$ 1.0, {[Ba/Eu]} $>$ 0                & {[Ba/Fe]} $>$ 1.0, {[Eu/Fe]} $>$ 1, {[Ba/Eu]} $>$ 0 & {\citep{2016A&A...587A..50A}} \\
{[Ba/Fe]} $>$ 0, $-0.5$ $<$ {[Sr/Ba]}  $<$ 0.8 & {[Ba/Fe]} $>$ 0, $-1.5$ $<$ {[Sr/Ba]}  $<$ $-0.5$ & {\citep{2019A&A...623A.128H}} \\
{[La/Eu]} $>$ 0.5 & 0 $<$ {[La/Eu]} $<$ 0.5 & {\citep{2021A&A...645A..61K}} \\
{[Ba/Fe]} $>$ 1.0 and                                          & {[Ba/Fe]} $>$ 1.0, {[Eu/Fe]} $>$ 1 and & {\citep{2021A&A...649A..49G}} \\
 {[Eu/Fe]} $<$ 1, {[Ba/Eu]} $>$ 0  and/or {[La/Eu]} $>$ 0.5 & 0 $<$ {[Ba/Eu]} $<$ 1  and/or 0.0 $<$ {[La/Eu]} $<$0.7 & \\
or {[Eu/Fe]} $>$ 1.0, {[Ba/Eu]} $>$ 1.0 and/or {[La/Eu]} $>$ 0.7 &   & \\
\hline      
\end{tabular}\\
\end{table*}

Different studies may classify the same star as a CEMP-s or a CEMP-rs star. For one of the most extensively studied CEMP stars, namely, HD~196944 \citep{1998A&A...337..216Z,2001Natur.412..793V,2002ApJ...580.1149A,2005A&A...440..321J,2010A&A...509A..93M,2014AJ....147..136R,2015ApJ...812..109P,2021A&A...645A..61K,2025AJ....169..172M,2025ApJ...995....2R}, its chemical composition has been investigated in numerous papers using high-resolution and high signal-to-noise (S/N) spectra obtained with many spectrographs/telescopes, including the UVES/VLT, CASPEC and HARPS/3.6 m 
 ESO telescope, HERMES/1.2 m Mercator telescope, HDS/Subaru, MIKE/Magellan, STIS/HST, and 2dCoude echelle spectrograph/2.7 m Harlan J. Smith Telescope at McDonald Observatory. Despite the numerous observations available, the origin of the neutron-capture element overabundances in HD~196944 remains unclear, and there is no consensus on whether this star should be classified as CEMP-s \citep{2015A&A...581A..22A,2025ApJ...995....2R} or CEMP-rs \citep{2021A&A...645A..61K,2025A&A...704A.103S}. This disagreement is not surprising, as different studies report significantly different [Ba/Eu]
\footnote{We use a standard designation, [X/Y] = $\log($N$_{\rm X}$/N$_{\rm Y}$)$_{*} - \log($N$_{\rm X}$/N$_{\rm Y}$)$_{\odot}$, where N$_{\rm X}$ and N$_{\rm Y}$ are total number densities of elements X and Y, respectively.} abundance ratios for HD~196944, ranging from 1.10 \citep{2025ApJ...995....2R} to 0.24 \citep{2021A&A...645A..61K}. Table~\ref{hd196944} summarizes the stellar atmosphere parameters, as well as the [Ba/Fe], [Eu/Fe], and [Ba/Eu] abundance ratios derived by different authors. In addition to differences in observational data and adopted stellar parameters, abundance determinations are also affected by different treatments of line formation, whether under the local thermodynamic equilibrium (LTE) assumption or taking departures from LTE (i.e., NLTE effects) into account. NLTE leads to higher Eu abundance but lower Ba abundances from the Ba\ii\ subordinate lines, resulting in a shift in [Ba/Eu] from 0.84 in LTE to 0.39 in NLTE in the case of HD~196944 \citep{2025A&A...704A.103S}.

\begin{table*}
\caption{Literature data on HD~196944.}        
\label{hd196944} 
\setlength{\tabcolsep}{1.mm}  
\footnotesize        
\centering          
\begin{tabular}{llllllllll}   
\hline      
\teff/log~g/\vt & {[Fe/H]} & {[Ba/H]} & {[Ba/Fe]} & {[Eu/H]}& {[Eu/Fe]} & {[Ba/Eu]} & {note}          & {class.} & {ref.}  \\
 \hline                                                           
 5390/2.13/2.0   & --2.21 & --1.36      & 0.85 &     --1.75 &    0.46    &  0.39  &  NLTE             & rs & {\citep{2025A&A...704A.103S}} \\ 
                 & --2.21 & --1.03      & 1.18 &     --1.87 &    0.34    &  0.84  &  LTE              & rs & {\citep{2025A&A...704A.103S}} \\ 
 5428/2.11/1.7   & --2.09 & --0.86      & 1.23 &     --1.96 &    0.13    &  1.10  &  LTE               & s & {\citep{2025ApJ...995....2R}} \\       
 5158/1.28/1.7   & --2.50 & --1.48      & 1.02 &     --2.00 &    0.50    &  0.52  &  LTE               & rs & {\citep{2021A&A...645A..61K}}  \\    
                 & --2.50 & -           & -    &     --1.72 &    0.78    &  0.24  &  NLTE: Eu, LTE: Ba & rs & {\citep{2021A&A...645A..61K}} \\
 5250/1.80/1.7   & --2.25 & --1.15      & 1.10 &     --2.08 &    0.17    &  0.93  &  LTE              & s & {\citep{2002ApJ...580.1149A}} \\ 
 5353/1.70/1.9   & --2.23 & --1.09      & 1.14 &     -      &    -       &  -     &   -               & & {\citep{2005A&A...440..321J}} \\     
 5250/1.70/1.9   & --2.45 & --0.89      & 1.56 &     -      &    -       &  -     &   -               & & {\citep{1998A&A...337..216Z}} \\  
\hline      
\end{tabular}
\end{table*}

A more robust classification method for CEMP-s and CEMP-rs stars, which uses additional heavy elements beyond Ba and Eu, has been proposed by \citet{2021A&A...645A..61K}. This approach is based on a comparison of stellar abundance patterns with the solar scaled r-process pattern. This method implies that CEMP-rs stars match the solar scaled r-process pattern better than CEMP-s stars.

In the search for a more physically motivated, rather than phenomenological, approach to distinguishing i-process and s-process tracers, \citet{2024Galax..12...89V} suggested using the Ba isotope ratio. In stars, Ba can be represented by five isotopes: $^{\rm 134}$Ba, $^{\rm 135}$Ba, $^{\rm 136}$Ba, $^{\rm 137}$Ba, and $^{\rm 138}$Ba. The isotopes $^{\rm 134}$Ba and $^{\rm 136}$Ba are s-only isotopes and cannot be produced in r- and i-processes due to the presence of stable r-only isotopes $^{\rm 134}$Xe and $^{\rm 136}$Xe, which block the $\beta$-decay pathway and prevent the formation of $^{\rm 134}$Ba and $^{\rm 136}$Ba. Therefore, different fractions of odd Ba isotopes (\fodd\ = (N($^{135}$Ba) + N($^{137}$Ba))/N(Ba)) are predicted for the s-process: \fodd\ = 0.10 \citep{2020MNRAS.491.1832P} and for the i-process: F$_{\rm odd}$ = 0.60 to 0.80 \citep{2025EPJA...61...68C}.
This idea is clear and reasonable, and the only limitation lies in methodological challenges arising from strong saturation of the Ba\ii\ resonance lines and blending with molecular carbon lines in CEMP stars, which obstructs accurate analysis of the Ba\ii\ resonance lines. 

For determination of the Ba isotope ratio, CEMP stars with high \teff\ and log~g are better suited than cool giants, as molecular carbon lines are weak or absent and the Ba\ii\ resonance lines are much less saturated. \citet{2025A&A...704A.103S} determined \fodd\ in nine hot CEMP-s and CEMP-rs stars with the Ba\ii\ resonance lines equivalent widths (EW) of less than 185~m\AA. It has been shown that, indeed, different CEMP stars have distinct Ba isotope ratios. 

The number of stars available for this kind of analysis remains very small, and, unfortunately, \citet{2025A&A...704A.103S} overlooked three CEMP stars that have Ba\ii\ resonance lines suitable for \fodd\ determinations. These stars are SDSS~J1349-
0229 and SDSS~J1036+
1212, which were classified as CEMP-rs and CEMP-s, respectively, by \citet{2010A&A...513A..72B}, and BPS~CS~29512-073, which was classified as CEMP-s by \citet{2012A&A...548A..34A}. All three stars were recently analyzed by \citet{2026ApJ...997...44R}, who confirmed their classification from the literature. In this study we aim to determine \fodd\ in these three stars and to check whether the derived values are consistent with the literature classification. 

The paper is structured as follows. In Section~\ref{sample_obs} we describe our sample stars, observations, and stellar atmosphere parameters. The abundance determination method is presented in Section~\ref{abund}. We discuss our findings in Section~\ref{results}, and we summarize our conclusions in Section~\ref{conclusions}.

\section{Stellar Sample, Observations, and Atmospheric Parameters}\label{sample_obs}

All three stars were analyzed by \citet{2026ApJ...997...44R}, who classified SDSS~J1349-0229 as a CEMP-rs star and BPS~CS~29512-073 and SDSS~J1036+1212 as CEMP-s stars by determining their detailed chemical abundance patterns. We checked all the 17 stars from \citet{2026ApJ...997...44R} and found that only three stars meet our selection criteria for the Ba\ii\ resonance lines not being too saturated for Ba isotope ratio analysis. 

We used high-resolution spectra from the UV-Visual Echelle Spectrograph (UVES) at the UT2 Kueyen Telescope that are available in the European Southern Observatory Science Archive\footnote{\url{http://archive.eso.org/wdb/wdb/adp/phase3\_main/form} (accessed on 30 April 2026).
}. 
The quality of the adopted spectra is reasonable, with a spectral resolving power of $\lambda/\Delta \lambda >$ 30,000 and a S/N per pixel of $>$30. The program IDs and the characteristics of the observed spectra are listed in Table~\ref{sample}. 

Stellar atmosphere parameters were determined using Gaia photometry and parallaxes, as well as the NLTE abundances from  Fe\ione\ and Fe\ii\ lines.  
We calculated the effective temperatures (\teff ) using the {\it Gaia
} ${\rm BP-G}$, ${\rm G-RP}$, ${\rm BP-RP}$ dereddened colors and the calibration of \citet{2021A&A...653A..90M}. The extinction ${\rm E(B-V)}$ was adopted from \citet{2019ApJ...887...93G}, and the colors were corrected according to \citet{2018MNRAS.479L.102C}. Different colors yield very similar effective temperatures, and the uncertainty in \teff\ is therefore mainly defined by an uncertainty in the calibration  of 80~K as given by \citet{2021A&A...653A..90M}.

\begin{table*}
\caption{Stellar sample, atmospheric parameters, and characteristics of the observed spectra.} 
\label{sample} 
\footnotesize        
\centering          
\begin{tabular}{lllllllll}   
\hline      
Name, type   &  \tiny{E(B-V)} & \teff ,  & log g$^*$, & \tiny{[Fe/H]}        & \vt , &  \tiny{S/N$_{\rm red}$} & R,& Program ID \\
         & & \tiny{K} & \tiny{$\rm cm~s^{-2}$} &     & \tiny{\kms} &       & 10$^3$ &         \\
\hline                                                           
 SDSS J1349-0229, rs & 0.05  & 6150 & 4.28 (0.12)          & --2.84 (0.10) & 1.2 &  70  & 31 & \tiny{078.D-0217(A), 383.D-0927(A) }  \\    
 SDSS J1036+1212, s  & 0.02 & 5720 & 3.47 (0.10)           & --3.27 (0.12) & 1.3 &  62  & 31 & \tiny{078.D-0217(A)}  \\
 BPS CS 29512-073, s  & 0.04 & 5760 & 3.65 (0.10)           & --1.88 (0.03) & 1.3 & 125  & 42 & \tiny{076.D-0451(A) }  \\     
\hline      
\end{tabular}
{\it Note.} * -- The log~g and [Fe/H] uncertainties are indicated in parentheses. For each sample star, the uncertainties in \teff\ and \vt\ of 80~K and 0.1~\kms, respectively, were adopted.    
\end{table*}

We calculated the surface gravities (log~g) using distances based on {\it Gaia} parallaxes. The parallaxes were corrected for the zero offset according to \citet{2021A&A...649A...4L}, and the distances were computed from the maximum of the probability distribution function as described by \citet{2015PASP..127..994B}. With these distances, effective temperatures, bolometric corrections of \citet{2018MNRAS.479L.102C}, and assuming a mass of 0.8 solar masses, we derived the surface gravities using the formula $\log g = 4.44+\log(m/m_{\odot})+0.4(M_{\rm bol} - 4.75) + 4\log($\teff$/5780)$, where $m_{\odot}$ is the solar mass, and $M_{\rm bol}$ is the absolute bolometric magnitude. 

The most metal-rich sample star \cs\ has the observed spectrum with the highest S/N ratio compared to the other two sample stars. Therefore, in \cs \ we determined iron abundances from 49 lines of Fe\ione\ and 8 lines of Fe\ii . For the lines of Fe\ione , we applied the NLTE abundance corrections from \citet{mash_fe}, which are available in the INASAN database\footnote{\url{https://spectrum.inasan.ru/nLTE2/}.} \citep{2023MNRAS.524.3526M}. The NLTE effects for the Fe\ii\ lines are negligible in the stellar parameter range we considered \citep{mash_fe,2019A&A...631A..43M}. When applying a 50~K lower \teff\ and 0.1~dex lower log~g compared to those derived for \cs\ from photometry, we found consistent within the error bars NLTE abundances from Fe\ione\ and Fe\ii , with the average values of $\eps$ = 5.53\footnote{We use a standard  abundance designation, $\eps$(X) = $\log($N$_{\rm X}$/N$_{\rm H}$) + 12, where N$_{\rm X}$ and N$_{\rm H}$ are total number densities of element X and hydrogen, respectively.} $\pm$ 0.08 and 5.57 $\pm$ 0.03, respectively. These stellar parameters are presented in Table~\ref{sample}. Iron abundances from individual lines with their atomic data are given in Table~\ref{fe_lines}. 

When applying the same list of iron lines to the two other sample stars, we detected only five lines of Fe\ione\ and Fe\ii\ (Table~\ref{fe_lines}). In \sjten , we found $\eps$ (Fe\ione ) = 4.33 $\pm$ 0.11 and $\eps$ (Fe\ii ) = 4.18 $\pm$ 0.10 in NLTE, when using photometric \teff\ and log~g. In \sj , we applied 70~K lower \teff\ and 0.12~dex higher log~g compared to those derived from photometry, and found in NLTE $\eps$ (Fe\ione ) = 4.74 $\pm$ 0.07 and $\eps$ (Fe\ii ) = 4.61 $\pm$ 0.10. In these two stars, the small number of iron lines and their small EWs prevented us from further tuning the stellar parameters for obtaining a perfect match between NLTE abundances from Fe\ione\ and Fe\ii. The microturbulent velocity was computed using an empirical relation based on NLTE analysis of Fe\ione\ and Fe\ii\ lines in dwarfs \citep{2015ApJ...808..148S}. 

We compared our stellar atmosphere parameters with those derived in the original studies of the sample stars. \citet{2010A&A...513A..72B} determined the effective temperatures from the wings of the H$\alpha$ line and the Fe\ione\ excitation equilibrium. The surface gravity was derived from the Fe\ione /Fe\ii\ ionization equilibrium. For \sj\ they found \teff /log~g = 6200~K/4.0. \teff\  is consistent within the error bars with our determinations of 6150~K $\pm$ 80~K, while we found a higher log~g = 4.28 $\pm$ 0.12. 
For \sjten\ \citet{2010A&A...513A..72B} determined \teff /log~g = 6000~K/4.0, while we found significantly lower \teff\ = 5720~K $\pm$ 80~K and log~g = 3.47 $\pm$ 0.10. It is known that the effective temperatures derived from the Fe\ione\ excitation equilibrium are systematically, by up to several hundred K, lower compared to those derived from photometry \citep{2010A&A...512A..54C,2013ApJ...769...57F} and underestimated \teff\  results in an underestimated log~g when determined from the Fe\ione /Fe\ii\ ionization equilibrium. For \sjten\, the situation is the opposite and it can hardly be understood. We attribute this discrepancy to a large uncertainty in \citet{2010A&A...513A..72B} determinations caused by evaluating stellar parameters from a small number of weak iron lines in \sjten.

While an accurate analysis of the Balmer line wings is beyond the scope of the present study, for illustration, we plot the H$\alpha$ line profiles in the sample stars together with their 1D~LTE synthetic spectra computed with \teff /log~g/[Fe/H] = 5720~K/3.47/$-3.3$ and 6150~K/4.28/$-2.8$ and representing \sjten\ and \sj , respectively (Figure~\ref{fit_halpha}). Figure~\ref{fit_halpha} demonstrates that \sjten\ is significantly cooler than \sj , and its temperature is very close to that of \cs , in line with our determination of \teff ~= 5720~K $\pm$ 80~K.

For \cs , \citet{2012A&A...548A..34A} determined \teff /log~g = 5560~K $\pm$ 50~K/3.44 $\pm$ 0.07 using an iterative process that employs photometric \teff\ and log~g from isochrones as a first step, and then revised them using Fe\ione\ excitation equilibrium and Fe\ione /Fe\ii\ ionization balance in LTE. We found \teff ~= 5760~K $\pm$ 80~K and log~g = 3.65 $\pm$ 0.10, supporting the idea that the effective temperatures derived from the Fe\ione\ excitation equilibrium are systematically lower compared to those derived from photometry. The H$\alpha$ line wings also support the adopted \teff\ (Figure~\ref{fit_halpha}).

\section{Abundance Analysis}\label{abund}

The line list was extracted from the Vienna Atomic Line Database  \vald\ \citep{1995A&AS..112..525P, 2019ARep...63.1010P}, which includes isotopic and hyperfine splitting (HFS). 
For Ba\ii\ lines, \vald\ provides the oscillator strengths (log~gf) from \citet{Miles_Wiese}. We compared these log~gf with the most recent laboratory results of \citet{2015PhRvA..91d0501D} for the Ba\ii\ 4934~\AA\ and 6496~\AA\ and \citet{2016NatSR...629772D} for the Ba\ii\ 4554~\AA , 5853~\AA , and 6141~\AA , where the branching factors were measured with an accuracy of better than one percent. The differences in log~gf values between the recent measurements and those from \citet{Miles_Wiese} amounts to 0.00, $-0.02$, $-0.02$, 0.01, and 0.01 for the Ba\ii\ 4554~\AA , 4934~\AA , 5853~\AA , 6141~\AA , and 6496~\AA , respectively. In this study, we used the oscillator strengths for Ba\ii\ lines from \citet{2015PhRvA..91d0501D} and \citet{2016NatSR...629772D}. For consistency, for our comparison sample stars from \citet{2025A&A...704A.103S}, we recomputed barium abundances and \fodd\  using the same log~gf values. 
The oscillator strengths for Eu\ii\ lines were taken from the laboratory measurements of \citet{LWHS}. The HFS components for the Ba\ii\ lines were taken from \citet{WABM} and for the Eu\ii\ lines from  \citet{HMAP}, \citet{MHAP}, and \citet{RESM}. We present the list of isotopic and HFS components of the Ba\ii\ and Eu\ii\ lines in Tables~\ref{ba_components} and~\ref{eu_components}, respectively. We provide a tool for generating the lists of Ba\ii\ lines for a given \fodd\ and it is available on GitHub\footnote{\url{https://github.com/sitamih/ba\_linelist}.}. 

The abundance for each spectral line is determined by fitting the synthetic line profile to the observed profile using the code \SV\ \cite{2019ASPC..518..247T}. A conjunction of the code \SV\ with the code BinMag \citep{2018ascl.soft05015K} enables the implementation of pre-computed departure coefficients, which represent the ratio of NLTE to LTE atomic level populations. The NLTE calculations are performed with a modified version of the \textsc{DETAIL} code \citep{detail}, where the opacity package was updated as described by \citet{mash_fe}. We adopted the Ba\ii\ model atom of  \citet{2019AstL...45..341M}, and the Eu\ii\ model atom of \citet{2000AA...364..249M}, which was updated by including data on inelastic collisions with hydrogen atoms from \citet{2024A&A...683A.200S}. Classical 1D model atmospheres from the \textsc{marcs} model grid \citep{marcs}, interpolated for given \teff, log~g, and [Fe/H] of the stars, were used.

The method of the Ba isotope ratio determination relies on the fact that the odd isotopes are subject to HFS of the energy levels, and a higher \fodd\ results in a broader line profile and greater total absorbed energy. HFS primarily affects the ground state, meaning that the Ba\ii\ resonance lines can serve as a diagnostic of the Ba isotope ratio. In contrast, subordinate lines are almost unaffected by the adopted isotope ratio and can be used as reliable indicators of barium abundance. This feature can be used for an \fodd\ determination by comparing abundances from the subordinate lines and the resonance lines computed with different \fodd. This idea was first proposed by \citet{1989ApJ...346.1030C}, who pointed out the importance of taking HFS into account when determining Ba abundances. The abundance comparison method of the \fodd\ determination was applied to MP stars in the Milky Way \citep{1999A&A...343..519M,2006A&A...456..313M,2008A&A...478..529M,2019AstL...45..341M,2025A&A...699A.262S} and in the Sculptor dwarf spheroidal galaxy  \citep[][]{2015A&A...583A..67J}, as well as to CEMP-s and CEMP-rs stars \citep{2025A&A...704A.103S}.

\begin{figure}[H]

\includegraphics[width=4.6 cm]{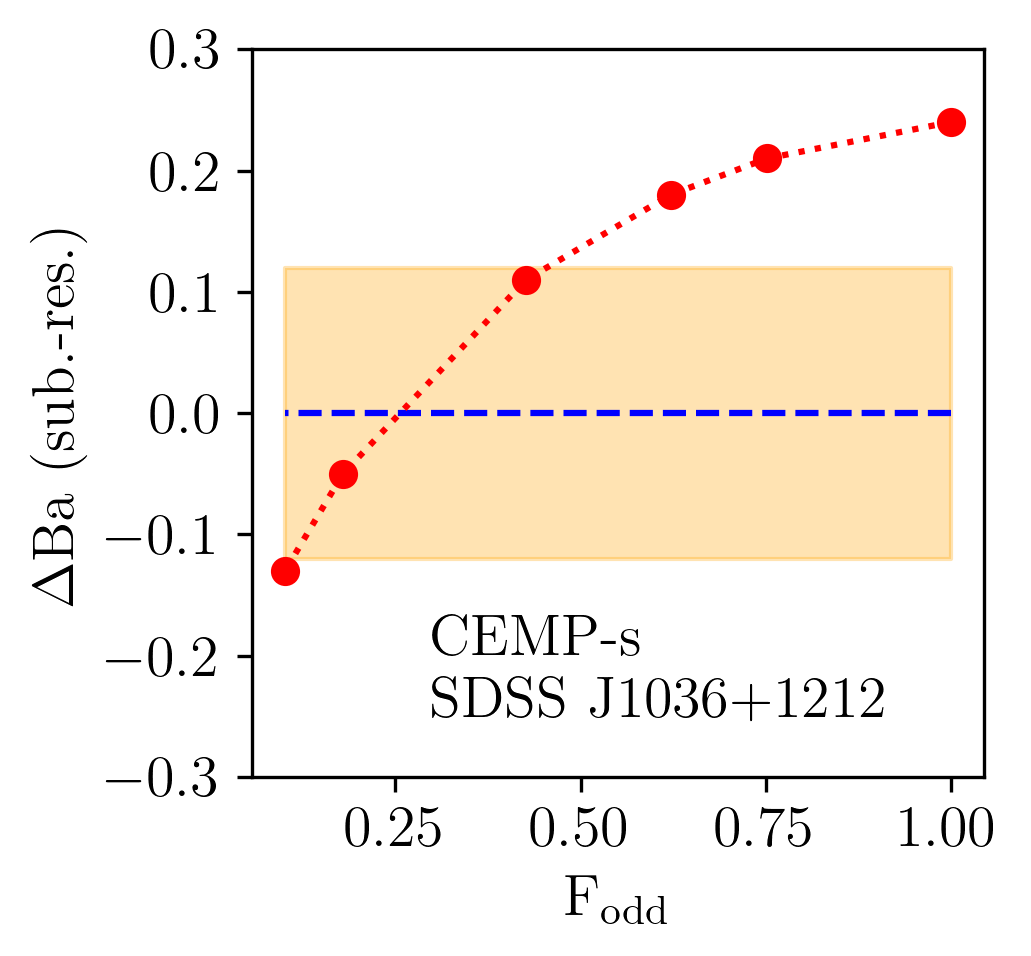}
\includegraphics[width=4.4 cm]{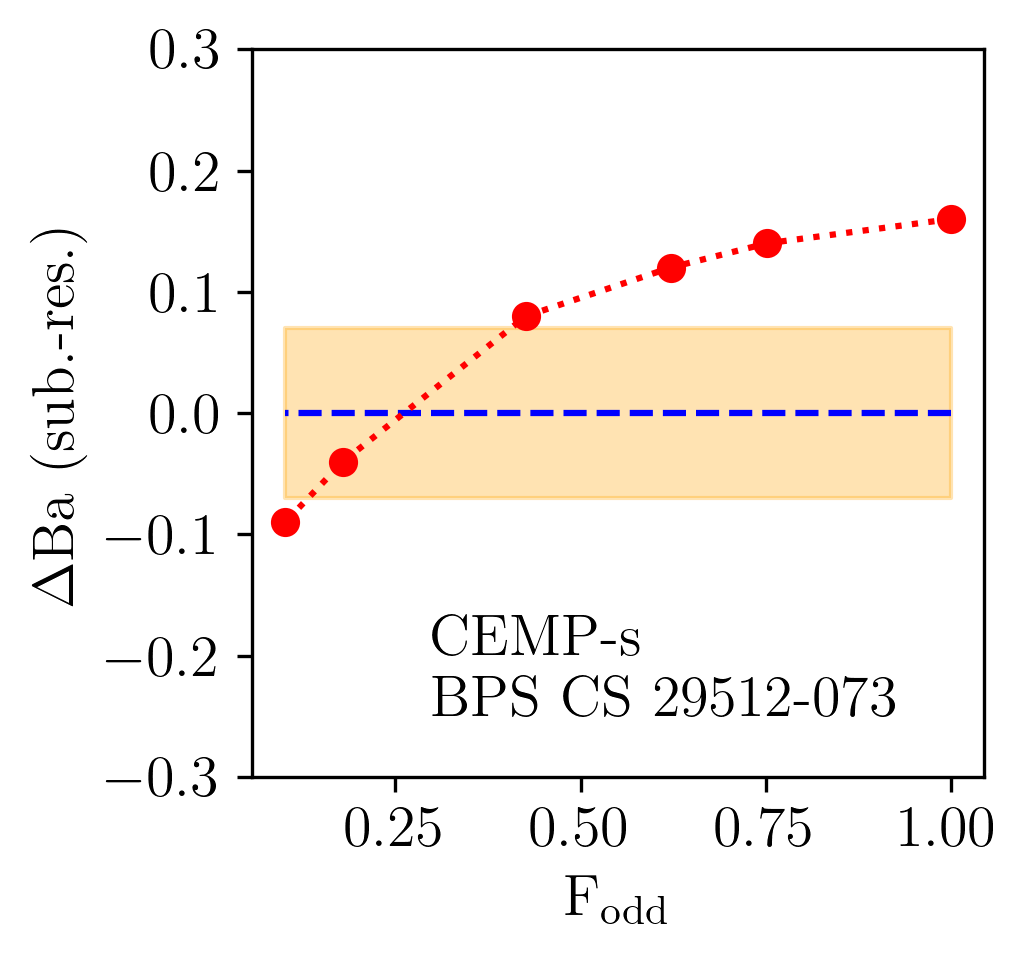}
\includegraphics[width=4.4 cm]{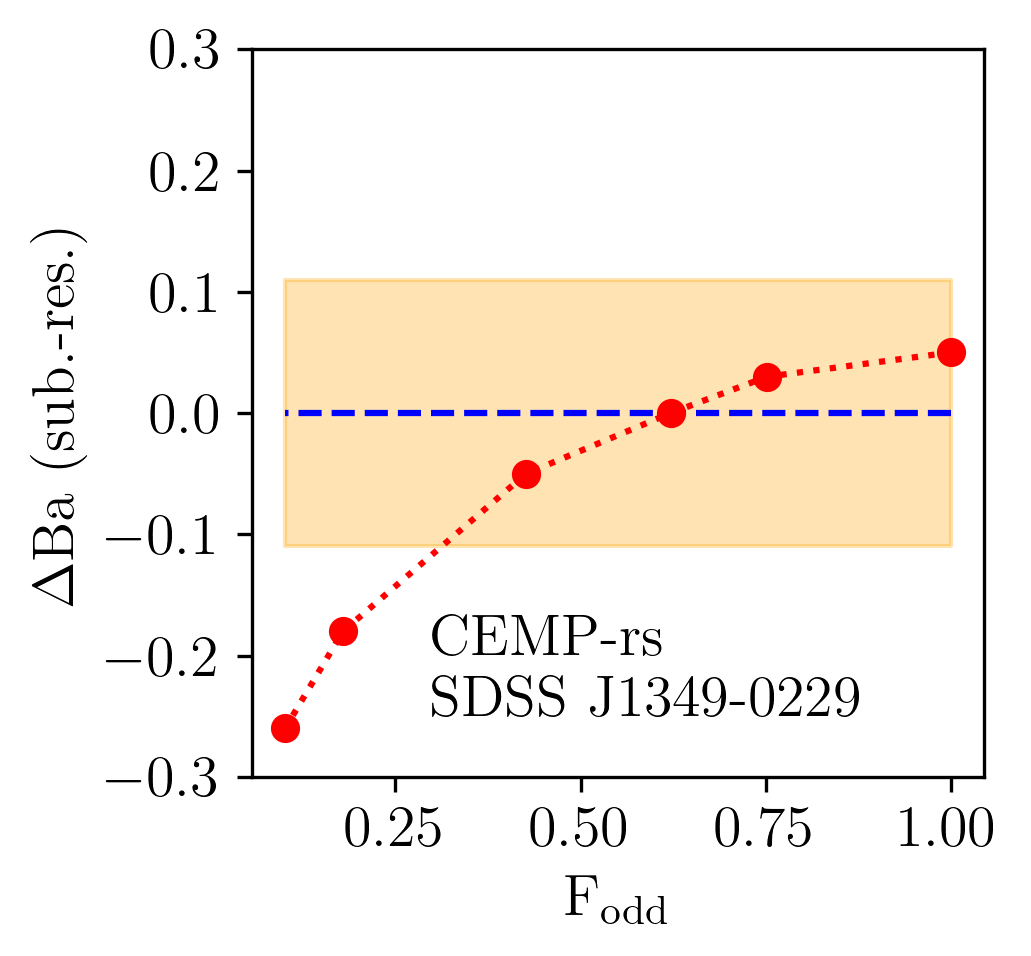}
\caption{NLTE 
 abundance difference between the subordinate and the resonance lines of Ba\ii\ as a function of \fodd\ in the sample stars (circles). Shaded area indicates the uncertainty in $\Delta$Ba(
sub.-res.). \label{ba_lines}}
\end{figure}

We analyzed four lines of Ba\ii: the strong resonance line at $\lambda$ = 4934~\AA, and the subordinate lines at $\lambda$ = 5853~\AA, 6141~\AA, and 6496~\AA. The resonance Ba\ii\  4554~\AA\ line is not covered by the available spectra of our sample stars. In our spectral synthesis calculations, we considered five Ba isotopes: $^{\rm 134}$Ba, $^{\rm 135}$Ba, $^{\rm 136}$Ba, $^{\rm 137}$Ba, and $^{\rm 138}$Ba. We determined the Ba abundances from the resonance line using different F$_{\rm odd}$: 0.10 (pure s-process), 0.18 (solar), 0.43, 0.62, 0.75 (pure r-process), and 1. Higher \fodd\ results in a stronger line and a lower derived barium abundance (Figure~\ref{ba_res_rs_same_abun}). When using \fodd\ = 0.1, we found 0.24~dex, 0.29~dex, and 0.34~dex higher abundances in \cs , \sj , and \sjten , respectively, compared to those derived with \fodd\ = 0.75 (Figure~\ref{ba_lines}). The abundances from individual lines of Ba\ii\ are presented in Table~\ref{lines}. 

The impact of \fodd\ on the Ba\ii\ 4934~\AA\ line profile might be less prominent compared to its impact on a total absorbed energy. In SDSS~J1349-0229 and SDSS~J1036+1212, EWs of the Ba\ii\ 4934~\AA\ lines  amount to 106~m\AA\ and 66~m\AA, respectively, while their best-fit spectra derived with different \fodd\  are almost indistinguishable (Figure~\ref{fit_ba4934}). In the CEMP-s star BPS~CS~29512-073 with a stronger Ba\ii\ 4934~\AA\ line with EW = 150~m\AA, the s-process Ba isotope mixture provides a better fit compared to those computed with the r-process mixture (Figure~\ref{fit_ba4934}).

We estimated an impact of 3D effects on the \fodd\ determination using the differences between 3D~NLTE and 1D~NLTE abundance corrections ($\Delta_{\rm 31}$) for the Ba\ii\ lines available at the ChETEC-INFRA project webpage\footnote{\url{https://www.chetec-infra.eu/3dnlte/abundance-corrections/barium/}.}. For a model atmosphere with \teff / log~g/ [Fe/H] / [Ba/Fe] = 5900~K/4/$-2$/1, the differences $\Delta_{\rm 31}$ amount to $-0.07$, $-0.02$, $-0.05$, and $-0.05$~dex for the Ba\ii\ 4934~\AA , 5853~\AA , 6141~\AA , and 6496~\AA , respectively. The lower abundance from the resonance line compared to those from the subordinate lines results in a smaller \fodd. The corresponding shift of 0.03~dex in $\Delta$Ba(sub-res) results in the shifts in \fodd\ of  $-0.11$, $-0.04$, and $-0.03$ in \sj , \cs , and \sjten , respectively, which can be considered minor compared to the uncertainties in \fodd\ determinations. 

\begin{table*}
\caption{NLTE and LTE abundances and EWs (m\AA) of the Ba\ii\ lines in the sample stars.}       
\label{lines}   
\footnotesize   
\centering          
\begin{tabular}{lrrrrrrrrrr} 
\hline      
 Name & Ba\ii:    &  \multicolumn{6}{c}{4934} &  5853 &  6141  & 6496  \\
 &   &  \multicolumn{6}{c}{----------------------------------------------------------------------} &    &    &   \\
& & \small{F$_{\rm odd}$:} 1.0 & 0.75 & 0.62 & 0.43 & 0.18 & 0.10 & & &  \\
\hline                   
SDSS J1349-0229     & NLTE &    0.98 &    1.00 &    1.03 &    1.08 &    1.21 &    1.29 &    1.09 &    0.97 &    1.02 \\
SDSS J1349-0229     & LTE  &    1.08 &    1.10 &    1.13 &    1.18 &    1.31 &    1.39 &    1.11 &    1.18 &    1.22 \\
SDSS J1349-0229     & EW   &   106.2 &   106.2 &   106.2 &   106.2 &   106.2 &   106.2 &    24.2 &    66.6 &    57.2 \\ 
BPS CS 29512-073    & NLTE &    1.24 &    1.25 &    1.27 &    1.32 &    1.43 &    1.49 &    1.40 &    1.23$^1$ &    1.39 \\
BPS CS 29512-073    & LTE  &    1.31 &    1.32 &    1.34 &    1.39 &    1.50 &    1.56 &    1.58 &    1.59$^1$ &    1.77 \\
BPS CS 29512-073    & EW   &   150.2 &   150.2 &   150.2 &   150.2 &   150.2 &   150.2 &    63.8 &   105.6 &   107.1 \\ 
SDSS J1036+1212     & NLTE &  --0.36 &  --0.33 &  --0.30 &  --0.23 &  --0.07 &    0.01 &      -- &  --0.19 &  --0.05 \\ 
SDSS J1036+1212     & LTE  &  --0.38 &  --0.35 &  --0.32 &  --0.25 &  --0.09 &  --0.01 &      -- &  --0.26 &  --0.10 \\ 
SDSS J1036+1212     & EW   &    66.3 &    66.3 &    66.3 &    66.3 &    66.3 &    66.3 &      -- &    28.8 &    28.3 \\ 
\hline      
\end{tabular}
{\it Note.} $^1$ - the Ba\ii\ 6141 \AA\ in \cs\ was excluded from the abundance analysis due to a significant contribution from the Fe\ione\ 6141~\AA\ blending line, see Fig.~\ref{fit_ba_sub}. 
\end{table*}

We determined the Eu abundance in BPS~CS~29512-073 from the Eu\ii\ 3819~\AA\ , 4129~\AA , and 4205~\AA\ lines, which yielded in NLTE $\eps$ = $-0.92$, $-0.77$, and $-0.87$, respectively. 
Our best-fit synthetic spectra together with the observed spectrum are presented in Figure~\ref{fit_eu}. In the spectra of the two other stars, the lines of Eu\ii\ were not detected, and we estimated upper limits on their Eu abundances. 

For each sample star, we provide the average NLTE abundance ratios in Table~\ref{ratios}. We calculated the uncertainties in the abundance ratios and \fodd\ including the uncertainties in the stellar atmosphere parameters, continuum placement, as well as the dispersion of the individual line measurements; see \citet{2025A&A...704A.103S} for details.

\begin{table*}
\caption{NLTE abundance ratios and fractions of odd Ba isotopes in the sample stars}
\setlength{\tabcolsep}{1.0mm}
 \label{ratios}
 \footnotesize   
\centering   
   \begin{tabular}{llrlrrrrll}    
      \hline
Name & [Ba/H] &  \multicolumn{2}{c}{[Eu/H]} &  \multicolumn{2}{c}{[Ba/Eu]} & [Ba/Fe] & \multicolumn{2}{c}{[Eu/Fe]} &  F$_{\rm odd}$ \\ 
   \hline 
SDSS J1349-0229  & --1.14  (0.12) & $<$ & --1.99 (0.22) & $>$ &   0.85  (0.23) & 1.70  (0.15) &  $<$ & 0.85  (0.15) &  0.65$_{-0.34}^{+0.35}$ \\
BPS CS 29512-073 & --0.78  (0.09) &     & --1.36 (0.09) &     &   0.59  (0.09) & 1.10  (0.08) &      & 0.52  (0.08) &  0.23$_{-0.10}^{+0.19}$ \\
SDSS J1036+1212  & --2.28  (0.13) & $<$ & --2.14 (0.22) & $>$ & --0.15  (0.23) & 0.98  (0.16) &  $<$ & 1.13  (0.16) &  0.23$_{-0.11}^{+0.22}$ \\
 \hline
      \end{tabular}\\
{\it Note.} The solar abundances are taken from \citet{2021SSRv..217...44L}. The total uncertainties are given in parentheses. 
\end{table*}

\section{Results}\label{results}

A CEMP-rs star SDSS~J1349-0229 shows \fodd\ = 0.65$_{-0.34}^{+0.35}$. Although the uncertainties are large, the possibility of a pure s-process or solar Ba isotope mixture can be excluded, while the derived \fodd\ agrees within the error bars with the i-process model predictions \citep{2025EPJA...61...68C}. The derived \fodd\ aligns well with \fodd\ from 0.44 to 0.57 found in the CEMP-rs stars studied in \citet{2025A&A...704A.103S}.

In the CEMP-s stars, we found \fodd\ =  0.23$_{-0.10}^{+0.19}$ in \cs\  and 0.23$_{-0.11}^{+0.22}$ in \sjten, which are consistent within the error bars with the solar value \fodd\ = 0.18 \citep{2020MNRAS.491.1832P} and the \fodd\ values  found in the CEMP-s stars studied in \citet{2025A&A...704A.103S}. 
The derived \fodd\ values argue that a contribution  from other n-capture sources besides the s-process to the chemical composition of the CEMP-s sample stars cannot be excluded. For example, in the solar system matter a contribution of 12~\% is attributed to the r-process \citep{2020MNRAS.491.1832P}. Therefore, a similar  r-process contribution can be expected for our CEMP-s sample stars.

\begin{figure}[H]

\includegraphics[width=13.5 cm]{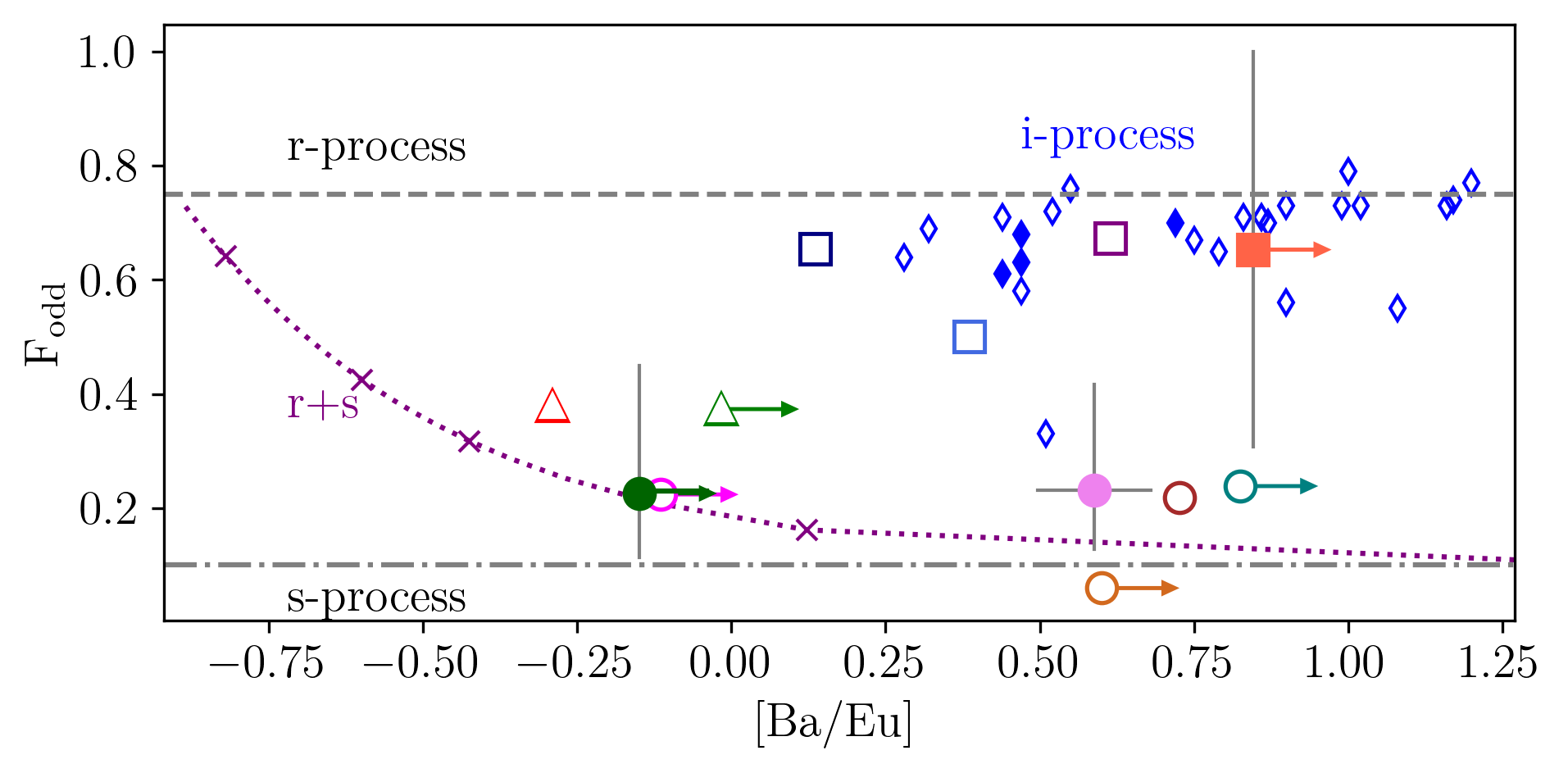}
\caption{F$_{\rm odd}$ 
 as a function of [Ba/Eu] in the sample stars:  SDSS~J1036+1212 (green circle), BPS~CS~29512-073 (rosy circle), and SDSS~J1349-0229 (orange square). The stars from \citet{2025A&A...704A.103S}\ are shown with open symbols. For comparison, we show the theoretical predictions for the r-process (dashed line) and the s-process (dash-dotted line). The s- and r-process material mixture is shown with the dotted line, where crosses mark relative r- to s-process material contributions of 5, 1, and 0.1 (from left to right). The i-process predictions are shown with diamonds, where the filled diamonds correspond to the i-process models with [Fe/H] and [Ba/Fe] that fall in the parameter range of our sample stars. \label{fodd_baeu}}
\end{figure}

Figure~\ref{fodd_baeu} shows \fodd\ as a function of [Ba/Eu] in the sample stars and our comparison sample from \citet{2025A&A...704A.103S}. We also plotted a curve that represents a mixture of the s-process and the r-process material. 
For illustration, we applied a pure s-process ratio of [Ba/Eu] = 1.25 as computed for the solar system material by \citet{2020MNRAS.491.1832P}. It is worth noting that individual models of AGB stars predict various [Ba/Eu] ratios. For example, \citet{2012ApJ...747....2L} found [Ba/Eu] from 0.57 to 1.09 for AGB stars with [Fe/H] = $-2.3$ and different masses.

Mixing  s- and r-process material in equal proportions yields [Ba/Eu] = $-0.6$ and \fodd\ = 0.4. This demonstrates that the chemical composition of the CEMP stars with \fodd\ $\simeq$ 0.6 and [Ba/Eu] $>$ 0 cannot be explained by a mixture of s- and r-process material and those stars are i-process tracers. In contrast, there is at least one star in our comparison sample stars that possibly gained its n-capture element overabundances from a mixture of the s- and r-process material in compatible proportions, resulting in [Ba/Eu] = $-0.29$ and \fodd\ =  0.38$_{-0.22}^{+0.53}$. This star is 2MASS~J09124370+0216236 and it could be considered a CEMP-r+s star (marked as a red open triangle in Figure~\ref{fodd_baeu}). 

\section{Conclusions}\label{conclusions}

We performed the NLTE abundance analysis of the Ba\ii\ lines in three CEMP stars and determined their \fodd. Our findings support the earlier classification of BPS~CS~29512-073 and SDSS~J1036+1212 as CEMP-s stars \citep{2012A&A...548A..34A,2010A&A...513A..72B,2026ApJ...997...44R} and of SDSS~J1349-0229 as a CEMP-rs star \citep{2010A&A...513A..72B,2026ApJ...997...44R}.
Our results are consistent with the finding of \citet{2025A&A...704A.103S}\ about distinct \fodd\ in CEMP-s and CEMP-rs stars. In total, we have determined \fodd\ in 12 CEMP stars, and based on their [Ba/Eu] and \fodd, we grouped them as follows: 

\begin{enumerate}
\item[--]
 CEMP-s 
 stars with [Ba/Eu] $> -0.1$ and \fodd\ $<$ 0.25. Heavy element abundances of these stars originate mainly from the s-process; however, a minor contribution from the r-process of less than 20\%  may still be present.

\item[--]CEMP-r + s stars with [Ba/Eu] $<$ 0 and \fodd\ $>$ 0.3. In these stars, their observed [Ba/Eu] and \fodd\ can be explained by a mixture of material produced by the r- and s-processes in compatible proportions of 1:2.

\item[--]CEMP-rs stars with [Ba/Eu] $>$ 0 and \fodd\ $\simeq$ 0.6. Heavy element abundances of these stars cannot be explained by a mixture of the material produced by the r- and s-processes, and they are the i-process tracers.
\end{enumerate}

Among the 12 CEMP stars analyzed in this study and \citet{2025A&A...704A.103S}, four stars are identified as i-process tracers. For obtaining observational constraints on i-process nucleosynthesis, further NLTE abundance determinations for many chemical elements are required.



\vspace{6pt} 

\authorcontributions{Both authors contributed to conceptualization, methodology, validation, and writing.} 

\funding{This research received no external funding.} 



\dataavailability{The data are available in tables and figures, as well as open sources listed in text.}  

\acknowledgments{
The authors thank the referees for providing valuable feedback.
}

\conflictsofinterest{The authors declare no conflict of interest.}  




%
%

\reftitle{References}


\bibliography{cemp_siren}

\appendixtitles{no} 
\appendixstart
\appendix

\section[\appendixname~\thesection]{}
\subsection[\appendixname~\thesubsection]{}

\begin{table*}
\caption{NLTE and LTE abundances from the Fe\ione\ and Fe\ii\ lines in the sample stars together with their atomic data and EWs}
\setlength{\tabcolsep}{1.0mm}
 \label{fe_lines}
 \footnotesize   
\centering   
   \begin{tabular}{llrrrrrrrrrrr}    
      \hline
Species & $\lambda$, &\eexc ,  &  log~gf &  $\eps$ &  $\eps$ & EW,  &  $\eps$ &  $\eps$ & EW,  &  $\eps$ &  $\eps$ & EW,  \\ 
          & \AA     & eV      &         &   NLTE  &   LTE  &  m\AA &   NLTE  &   LTE  &  m\AA &   NLTE  &   LTE  &  m\AA \\ 
          & \AA     & eV      &         &   \multicolumn{3}{c}{BPS CS 29512-073}  &   \multicolumn{3}{c}{SDSS J1036+1212} &  \multicolumn{3}{c}{SDSS J1349-0229} \\           
   \hline 
Fe\ione\ & 4891.49 & 2.85 & --0.11 &    5.57 &    5.55 & 82.5 &  4.44 &    4.27 & 23.9 & 4.71 &    4.61 & 24.6 \\ 
Fe\ione\ & 4903.31 & 2.88 & --0.93 &    5.51 &    5.47 & 36.5 &  &  &  &  &  &  \\ 
Fe\ione\ & 4918.99 & 2.87 & --0.34 &    5.51 &    5.48 & 66.9 &  &  &  &  &  &  \\ 
Fe\ione\ & 4938.81 & 2.88 & --1.08 &    5.46 &    5.42 & 27.7 &  &  &  &  &  &  \\ 
Fe\ione\ & 4966.09 & 3.33 & --0.89 &    5.55 &    5.51 & 21.1 &  &  &  &  &  &  \\ 
Fe\ione\ & 5001.86 & 3.88 &   0.01 &    5.39 &    5.34 & 29.1 &  &  &  &  &  &  \\ 
Fe\ione\ & 5049.82 & 2.28 & --1.36 &    5.57 &    5.53 & 46.7 &  &  &  &  &  &  \\ 
Fe\ione\ & 5068.77 & 2.94 & --1.04 &    5.54 &    5.50 & 30.6 &  &  &  &  &  &  \\ 
Fe\ione\ & 5074.75 & 4.22 & --0.20 &    5.71 &    5.66 & 20.6 &  &  &  &  &  &  \\ 
Fe\ione\ & 5171.60 & 1.48 & --1.75 &    5.58 &    5.54 & 64.8 &  &  &  & 4.83 &    4.75 & 16.9  \\ 
Fe\ione\ & 5191.45 & 3.04 & --0.55 &    5.49 &    5.45 & 47.8 &  &  &  &  &  &  \\ 
Fe\ione\ & 5192.34 & 3.00 & --0.52 &    5.55 &    5.51 & 54.1 &  &  &  &  &  &  \\ 
Fe\ione\ & 5194.94 & 1.56 & --2.09 &    5.57 &    5.53 & 46.6 &  &  &  &  &  &  \\ 
Fe\ione\ & 5215.18 & 3.27 & --0.93 &    5.50 &    5.46 & 20.7 &  &  &  &  &  &  \\ 
Fe\ione\ & 5216.27 & 1.61 & --2.10 &    5.54 &    5.50 & 42.6 &  &  &  &  &  &  \\ 
Fe\ione\ & 5217.39 & 3.21 & --1.07 &    5.44 &    5.41 & 16.3 &  &  &  &  &  &  \\ 
Fe\ione\ & 5232.94 & 2.94 & --0.07 &    5.44 &    5.42 & 74.9 &  4.22 &    4.04 & 15.3 & 4.67 &    4.57 & 22.1   \\ 
Fe\ione\ & 5266.55 & 3.00 & --0.39 &    5.45 &    5.41 & 55.5 &  &  &  &  &  &  \\ 
Fe\ione\ & 5281.79 & 3.04 & --0.83 &    5.49 &    5.45 & 34.0 &  &  &  &  &  &  \\ 
Fe\ione\ & 5283.62 & 3.24 & --0.52 &    5.54 &    5.49 & 41.3 &  &  &  &  &  &  \\ 
Fe\ione\ & 5302.30 & 3.28 & --0.88 &    5.59 &    5.55 & 25.9 &  &  &  &  &  &  \\ 
Fe\ione\ & 5307.36 & 1.61 & --2.99 &    5.59 &    5.55 & 11.2 &  &  &  &  &  &  \\ 
Fe\ione\ & 5324.18 & 3.21 & --0.10 &    5.45 &    5.41 & 60.1 &  &  &  &  &  &  \\ 
Fe\ione\ & 5339.93 & 3.27 & --0.68 &    5.51 &    5.47 & 31.7 &  &  &  &  &  &  \\ 
Fe\ione\ & 5367.47 & 4.42 &   0.55 &    5.34 &    5.28 & 28.1 &  &  &  &  &  &  \\ 
Fe\ione\ & 5369.96 & 4.37 &   0.54 &    5.40 &    5.34 & 32.4 &  &  &  &  &  &  \\ 
Fe\ione\ & 5383.37 & 4.31 &   0.50 &    5.54 &    5.48 & 39.6 &  &  &  &  &  &  \\ 
Fe\ione\ & 5393.17 & 3.24 & --0.71 &    5.46 &    5.42 & 29.3 &  &  &  &  &  &  \\ 
Fe\ione\ & 5400.50 & 4.37 & --0.15 &    5.54 &    5.49 & 12.8 &  &  &  &  &  &  \\ 
Fe\ione\ & 5415.20 & 4.39 &   0.51 &    5.51 &    5.45 & 35.4 &  &  &  &  &  &  \\ 
Fe\ione\ & 5424.07 & 4.32 &   0.52 &    5.57 &    5.51 & 42.1 &  &  &  &  &  &  \\ 
Fe\ione\ & 5569.62 & 3.42 & --0.54 &    5.51 &    5.46 & 31.5 &  &  &  &  &  &  \\ 
Fe\ione\ & 5572.84 & 3.40 & --0.31 &    5.46 &    5.41 & 41.0 &  &  &  &  &  &  \\ 
Fe\ione\ & 5576.09 & 3.43 & --1.00 &    5.67 &    5.63 & 19.3 &  &  &  &  &  &  \\ 
Fe\ione\ & 5586.76 & 3.37 & --0.14 &    5.47 &    5.42 & 51.5 &  &  &  &  &  &  \\ 
Fe\ione\ & 5615.64 & 3.33 &   0.05 &    5.43 &    5.39 & 61.9 &  &  &  &  &  &  \\ 
Fe\ione\ & 6024.06 & 4.55 & --0.11 &    5.69 &    5.65 & 14.1 &  &  &  &  &  &  \\ 
Fe\ione\ & 6136.61 & 2.45 & --1.50 &    5.62 &    5.58 & 37.3 &  &  &  &  &  &  \\ 
Fe\ione\ & 6137.69 & 2.59 & --1.37 &    5.53 &    5.49 & 32.9 &  &  &  &  &  &  \\ 
Fe\ione\ & 6191.56 & 2.43 & --1.42 &    5.48 &    5.44 & 35.3 &  &  &  &  &  &  \\ 
Fe\ione\ & 6230.72 & 2.56 & --1.28 &    5.56 &    5.52 & 40.1 &  &  &  &  &  &  \\ 
Fe\ione\ & 6252.55 & 2.40 & --1.76 &    5.63 &    5.59 & 28.2 &  &  &  &  &  &  \\ 
Fe\ione\ & 6301.50 & 3.65 & --0.72 &    5.53 &    5.48 & 17.1 &  &  &  &  &  &  \\ 
Fe\ione\ & 6335.33 & 2.20 & --2.23 &    5.58 &    5.54 & 16.9 &  &  &  &  &  &  \\ 
Fe\ione\ & 6393.60 & 2.43 & --1.43 &    5.42 &    5.38 & 32.9 &  &  &  &  &  &  \\ 
Fe\ione\ & 6400.00 & 3.60 & --0.52 &    5.65 &    5.60 & 32.0 &  &  &  &  &  &  \\ 
Fe\ione\ & 6421.35 & 2.28 & --2.01 &    5.59 &    5.55 & 22.0 &  &  &  &  &  &  \\ 
Fe\ione\ & 6430.84 & 2.18 & --1.95 &    5.57 &    5.53 & 28.0 &  &  &  &  &  &  \\ 
Fe\ione\ & 6494.98 & 2.40 & --1.27 &    5.57 &    5.53 & 48.9 &  &  &  &  &  &  \\ 
Fe\ii\   & 4491.40 & 2.86 & --2.65 &    5.57 &    5.57 & 23.7 &  &  &  &  &  &  \\ 
Fe\ii\   & 4508.28 & 2.86 & --2.23 &    5.55 &    5.55 & 40.9 &  &  &  &  &  &  \\ 
Fe\ii\   & 4923.92 & 2.89 & --1.39 &    5.62 &    5.62 & 79.8 & 4.30 &    4.30 & 26.6  &  &  &  \\ 
Fe\ii\   & 5018.44 & 2.89 & --1.23 &    5.62 &    5.62 & 87.7 & 4.17 &    4.17 & 28.4  &  4.51 &    4.51 & 27.1 \\ 
Fe\ii\   & 5169.03 & 2.89 & --1.14 &         &         &      & 4.06 &    4.06 & 27.6  &  4.70 &    4.70 & 39.5  \\ 
Fe\ii\   & 5197.57 & 3.23 & --2.24 &    5.57 &    5.57 & 27.0 &  &  &  &  &  &  \\ 
Fe\ii\   & 5234.62 & 3.22 & --2.17 &    5.55 &    5.55 & 29.5 &  &  &  &  &  &  \\ 
Fe\ii\   & 5276.00 & 3.20 & --2.10 &    5.51 &    5.51 & 31.9 &  &  &  &  &  &  \\ 
Fe\ii\   & 6456.38 & 3.90 & --2.07 &    5.58 &    5.58 & 12.6 &  &  &  &  &  &  \\   
 \hline
      \end{tabular}\\
\end{table*}

\begin{table*}
\caption{The HFS and isotopic components of the Ba\ii\ lines. Solar isotope mixture is adopted. }
 \label{ba_components}
 \footnotesize   
\centering   
   \begin{tabular}{llrr|llrr}    
      \hline
A  & $\lambda$, &\eexc ,  &  log~gf & A  & $\lambda$, &\eexc ,  &  log~gf   \\ 
   & \AA       & eV      &           &  & \AA       & eV      &           \\         
   \hline 
137 &        4553.9980 &    0.0000 & --1.586 & 137 &        6141.7086 &    0.7036 & --2.254 \\
137 &        4553.9991 &    0.0000 & --1.586 & 137 &        6141.7086 &    0.7036 & --1.446 \\
137 &        4553.9993 &    0.0000 & --1.984 & 137 &        6141.7094 &    0.7036 & --3.400 \\
135 &        4554.0009 &    0.0000 & --1.817 & 135 &        6141.7105 &    0.7036 & --1.677 \\
135 &        4554.0021 &    0.0000 & --1.817 & 135 &        6141.7107 &    0.7036 & --2.485 \\
135 &        4554.0025 &    0.0000 & --2.215 & 135 &        6141.7113 &    0.7036 & --3.631 \\
130 &        4554.0310 &    0.0000 & --2.805 & 138 &        6141.7129 &    0.7036 & --0.214 \\
132 &        4554.0315 &    0.0000 & --2.826 & 136 &        6141.7140 &    0.7036 & --1.175 \\
134 &        4554.0319 &    0.0000 & --1.447 & 137 &        6141.7146 &    0.7036 & --1.652 \\
136 &        4554.0323 &    0.0000 & --0.935 & 137 &        6141.7154 &    0.7036 & --2.157 \\
138 &        4554.0336 &    0.0000 &   0.026 & 134 &        6141.7154 &    0.7036 & --1.687 \\
135 &        4554.0479 &    0.0000 & --1.370 & 135 &        6141.7157 &    0.7036 & --1.883 \\
137 &        4554.0503 &    0.0000 & --1.139 & 137 &        6141.7163 &    0.7036 & --3.224 \\
135 &        4554.0506 &    0.0000 & --1.817 & 135 &        6141.7163 &    0.7036 & --2.388 \\
135 &        4554.0518 &    0.0000 & --2.516 & 132 &        6141.7169 &    0.7036 & --3.066 \\
137 &        4554.0536 &    0.0000 & --1.586 & 135 &        6141.7170 &    0.7036 & --3.455 \\
137 &        4554.0547 &    0.0000 & --2.285 & 137 &        6141.7174 &    0.7036 & --1.902 \\
137 &        4934.0298 &    0.0000 & --1.627 & 130 &        6141.7183 &    0.7036 & --3.045 \\
135 &        4934.0339 &    0.0000 & --1.858 & 137 &        6141.7183 &    0.7036 & --2.270 \\
137 &        4934.0419 &    0.0000 & --2.326 & 135 &        6141.7184 &    0.7036 & --2.133 \\
135 &        4934.0447 &    0.0000 & --2.557 & 137 &        6141.7188 &    0.7036 & --2.224 \\
130 &        4934.0744 &    0.0000 & --3.147 & 135 &        6141.7191 &    0.7036 & --2.501 \\
132 &        4934.0750 &    0.0000 & --3.168 & 135 &        6141.7198 &    0.7036 & --2.455 \\
134 &        4934.0755 &    0.0000 & --1.789 & 137 &        6496.8826 &    0.6043 & --2.821 \\
136 &        4934.0758 &    0.0000 & --1.277 & 135 &        6496.8857 &    0.6043 & --3.052 \\
138 &        4934.0773 &    0.0000 & --0.316 & 137 &        6496.8873 &    0.6043 & --2.122 \\
135 &        4934.0922 &    0.0000 & --1.858 & 135 &        6496.8901 &    0.6043 & --2.353 \\
137 &        4934.0950 &    0.0000 & --1.627 & 137 &        6496.8959 &    0.6043 & --1.675 \\
135 &        4934.1030 &    0.0000 & --1.858 & 135 &        6496.8977 &    0.6043 & --1.906 \\
137 &        4934.1071 &    0.0000 & --1.627 & 138 &        6496.8977 &    0.6043 & --0.510 \\
137 &        5853.6685 &    0.6043 & --3.031 & 136 &        6496.8988 &    0.6043 & --1.471 \\
137 &        5853.6701 &    0.6043 & --2.973 & 134 &        6496.9004 &    0.6043 & --1.983 \\
137 &        5853.6703 &    0.6043 & --3.177 & 137 &        6496.9015 &    0.6043 & --2.520 \\
135 &        5853.6705 &    0.6043 & --3.262 & 132 &        6496.9021 &    0.6043 & --3.362 \\
135 &        5853.6715 &    0.6043 & --3.204 & 135 &        6496.9025 &    0.6043 & --2.751 \\
135 &        5853.6718 &    0.6043 & --3.408 & 137 &        6496.9035 &    0.6043 & --2.122 \\
137 &        5853.6720 &    0.6043 & --3.575 & 130 &        6496.9037 &    0.6043 & --3.341 \\
137 &        5853.6723 &    0.6043 & --3.177 & 135 &        6496.9044 &    0.6043 & --2.353 \\
135 &        5853.6734 &    0.6043 & --3.806 & 137 &        6496.9082 &    0.6043 & --2.122 \\
137 &        5853.6739 &    0.6043 & --2.876 & 135 &        6496.9088 &    0.6043 & --2.353 \\
135 &        5853.6740 &    0.6043 & --3.408 &     &                  &           &         \\
138 &        5853.6742 &    0.6043 & --1.167 &     &                  &           &         \\
135 &        5853.6750 &    0.6043 & --3.107 &     &                  &           &         \\
136 &        5853.6751 &    0.6043 & --2.128 &     &                  &           &         \\
137 &        5853.6755 &    0.6043 & --2.429 &     &                  &           &         \\
137 &        5853.6758 &    0.6043 & --2.973 &     &                  &           &         \\
134 &        5853.6764 &    0.6043 & --2.640 &     &                  &           &         \\
135 &        5853.6766 &    0.6043 & --2.660 &     &                  &           &         \\
135 &        5853.6769 &    0.6043 & --3.204 &     &                  &           &         \\
137 &        5853.6809 &    0.6043 & --3.031 &     &                  &           &         \\
135 &        5853.6812 &    0.6043 & --3.262 &     &                  &           &         \\
 \hline
      \end{tabular}\\
{\it Note.} We provide a tool for generating the lists of Ba\ii\ lines for a given \fodd , which is available on https://github.com/sitamih/ba\_linelist.
\end{table*}

\begin{table*}
\caption{The HFS and isotopic components of the Eu\ii\ lines. Solar isotope mixture is adopted. }
\setlength{\tabcolsep}{1.0mm}
 \label{eu_components}
 \footnotesize   
\centering   
   \begin{tabular}{llrr|llrr|llrr}    
      \hline
A  & $\lambda$, &\eexc ,  &  log~gf & A  & $\lambda$, &\eexc ,  &  log~gf & A  & $\lambda$, &\eexc ,  &  log~gf   \\ 
   & \AA       & eV      &           &  & \AA       & eV      &            &  & \AA       & eV      &           \\         
   \hline 
  151  &   3819.57682 &  0.00 & --0.940 &  151  &   4129.59676 &  0.00 & --1.832  & 151  &   4204.89432 &  0.00 & --1.432  \\
  151  &   3819.59380 &  0.00 & --0.831 &  151  &   4129.60032 &  0.00 & --1.355  & 151  &   4204.89746 &  0.00 & --1.733  \\
  151  &   3819.59557 &  0.00 & --1.609 &  151  &   4129.61392 &  0.00 & --1.636  & 151  &   4204.90262 &  0.00 & --2.687  \\
  151  &   3819.61750 &  0.00 & --0.722 &  151  &   4129.61868 &  0.00 & --1.297  & 151  &   4204.92018 &  0.00 & --1.256  \\
  151  &   3819.62005 &  0.00 & --1.419 &  151  &   4129.62223 &  0.00 & --1.832  & 151  &   4204.92534 &  0.00 & --1.550  \\
  151  &   3819.62182 &  0.00 & --2.827 &  151  &   4129.63885 &  0.00 & --1.577  & 151  &   4204.93242 &  0.00 & --2.578  \\
  153  &   3819.64347 &  0.00 & --0.902 &  151  &   4129.64460 &  0.00 & --1.167  & 151  &   4204.95715 &  0.00 & --1.093  \\
  151  &   3819.64771 &  0.00 & --0.617 &  151  &   4129.64936 &  0.00 & --1.636  & 151  &   4204.96423 &  0.00 & --1.491  \\
  151  &   3819.65124 &  0.00 & --1.365 &  151  &   4129.67183 &  0.00 & --1.614  & 151  &   4204.97307 &  0.00 & --2.687  \\
  153  &   3819.65180 &  0.00 & --1.571 &  153  &   4129.67754 &  0.00 & --1.794  & 153  &   4204.99375 &  0.00 & --1.394  \\
  153  &   3819.65248 &  0.00 & --0.793 &  151  &   4129.67829 &  0.00 & --1.016  & 153  &   4204.99554 &  0.00 & --1.695  \\
  151  &   3819.65379 &  0.00 & --2.681 &  153  &   4129.68030 &  0.00 & --1.317  & 153  &   4204.99834 &  0.00 & --2.649  \\
  153  &   3819.66347 &  0.00 & --2.789 &  153  &   4129.68396 &  0.00 & --1.598  & 151  &   4205.00512 &  0.00 & --0.947  \\
  153  &   3819.66415 &  0.00 & --1.381 &  151  &   4129.68404 &  0.00 & --1.577  & 153  &   4205.00564 &  0.00 & --1.218  \\
  153  &   3819.66433 &  0.00 & --0.684 &  153  &   4129.68728 &  0.00 & --1.259  & 153  &   4205.00844 &  0.00 & --1.512  \\
  153  &   3819.67847 &  0.00 & --0.579 &  153  &   4129.69004 &  0.00 & --1.794  & 153  &   4205.01198 &  0.00 & --2.540  \\
  153  &   3819.67915 &  0.00 & --2.643 &  153  &   4129.69428 &  0.00 & --1.539  & 151  &   4205.01397 &  0.00 & --1.525  \\
  153  &   3819.67932 &  0.00 & --1.327 &  153  &   4129.69760 &  0.00 & --1.129  & 153  &   4205.02258 &  0.00 & --1.055  \\
  151  &   3819.68419 &  0.00 & --0.518 &  153  &   4129.70092 &  0.00 & --1.598  & 151  &   4205.02439 &  0.00 & --3.030  \\
  151  &   3819.68895 &  0.00 & --1.407 &  153  &   4129.70920 &  0.00 & --1.576  & 153  &   4205.02612 &  0.00 & --1.453  \\
  151  &   3819.69249 &  0.00 & --2.768 &  153  &   4129.71182 &  0.00 & --0.978  & 153  &   4205.03002 &  0.00 & --2.649  \\
  153  &   3819.69428 &  0.00 & --0.480 &  151  &   4129.71321 &  0.00 & --1.800  & 153  &   4205.04430 &  0.00 & --0.909  \\
  153  &   3819.69680 &  0.00 & --1.369 &  153  &   4129.71513 &  0.00 & --1.539  & 153  &   4205.04820 &  0.00 & --1.487  \\
  153  &   3819.69765 &  0.00 & --2.730 &  151  &   4129.72004 &  0.00 & --0.865  & 153  &   4205.05198 &  0.00 & --2.992  \\
  153  &   3819.71099 &  0.00 & --0.387 &  151  &   4129.72650 &  0.00 & --1.614  & 151  &   4205.06395 &  0.00 & --0.816  \\
  153  &   3819.71594 &  0.00 & --1.559 &  153  &   4129.72958 &  0.00 & --1.762  & 153  &   4205.07042 &  0.00 & --0.778  \\
  153  &   3819.71846 &  0.00 & --3.058 &  153  &   4129.73063 &  0.00 & --0.827  & 153  &   4205.07419 &  0.00 & --1.670  \\
  151  &   3819.72663 &  0.00 & --0.425 &  153  &   4129.73324 &  0.00 & --1.576  & 151  &   4205.07437 &  0.00 & --1.708  \\
  151  &   3819.73292 &  0.00 & --1.597 &  153  &   4129.75490 &  0.00 & --0.683  & 153  &   4205.10044 &  0.00 & --0.658  \\
  151  &   3819.73769 &  0.00 & --3.096 &  153  &   4129.75594 &  0.00 & --1.762  & 151  &   4205.13344 &  0.00 & --0.696  \\
       &              &       &         &  151  &   4129.77018 &  0.00 & --0.721  &      &              &       &          \\
       &              &       &         &  151  &   4129.77701 &  0.00 & --1.800  &      &              &       &          \\
 \hline
      \end{tabular}\\
\end{table*}

\begin{figure}[H]
\centering
\includegraphics[trim={0 10 0 0},width=\hsize]{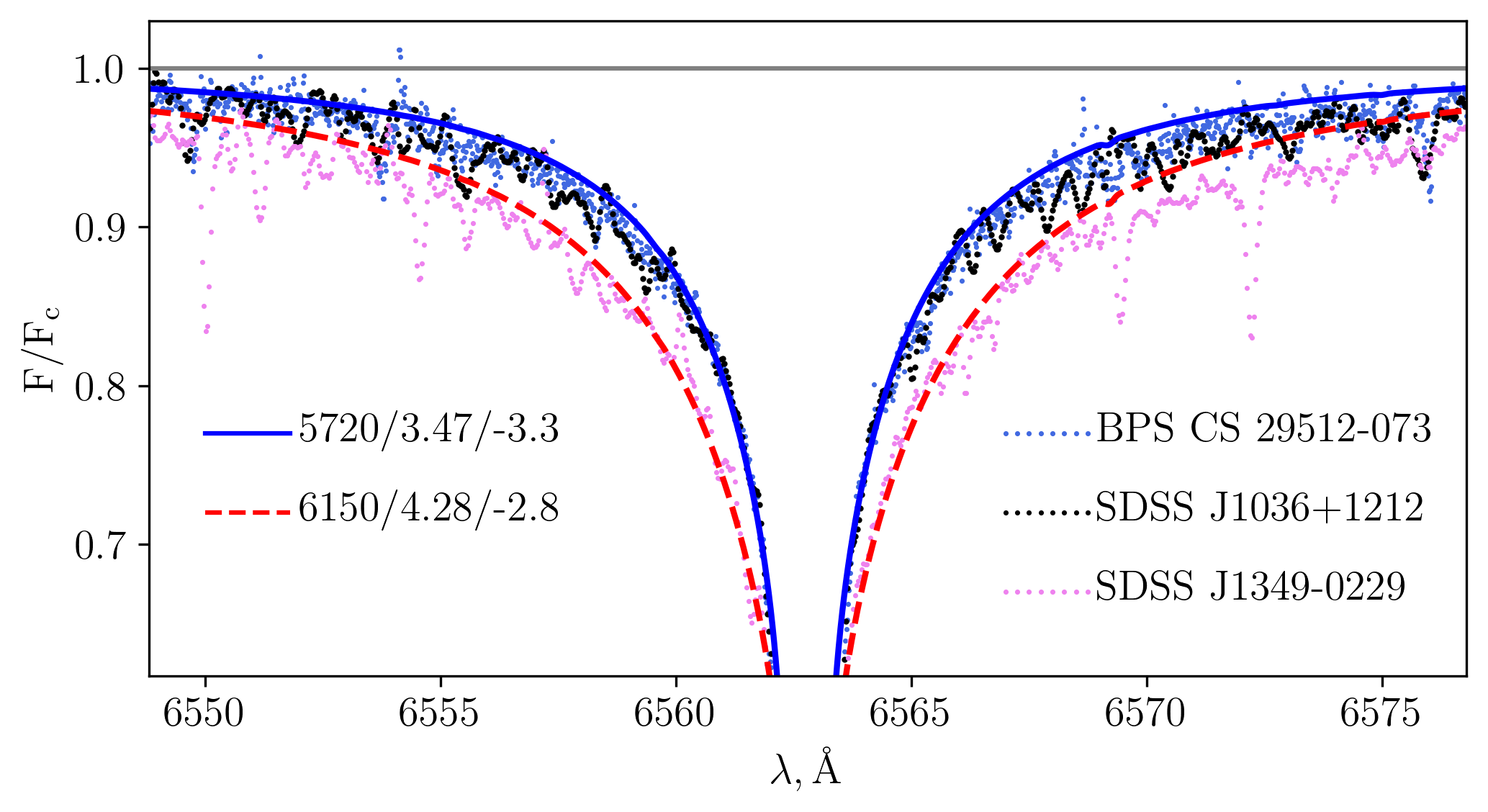}
\caption{The H$\alpha$ line profiles in the observed spectra (circles) of the sample stars along with the synthetic LTE spectra (lines). See legend for designations.
\label{fit_halpha}}
\end{figure}

\begin{figure}[H]
\centering
\includegraphics[trim={0 10 0 0},width=\hsize]{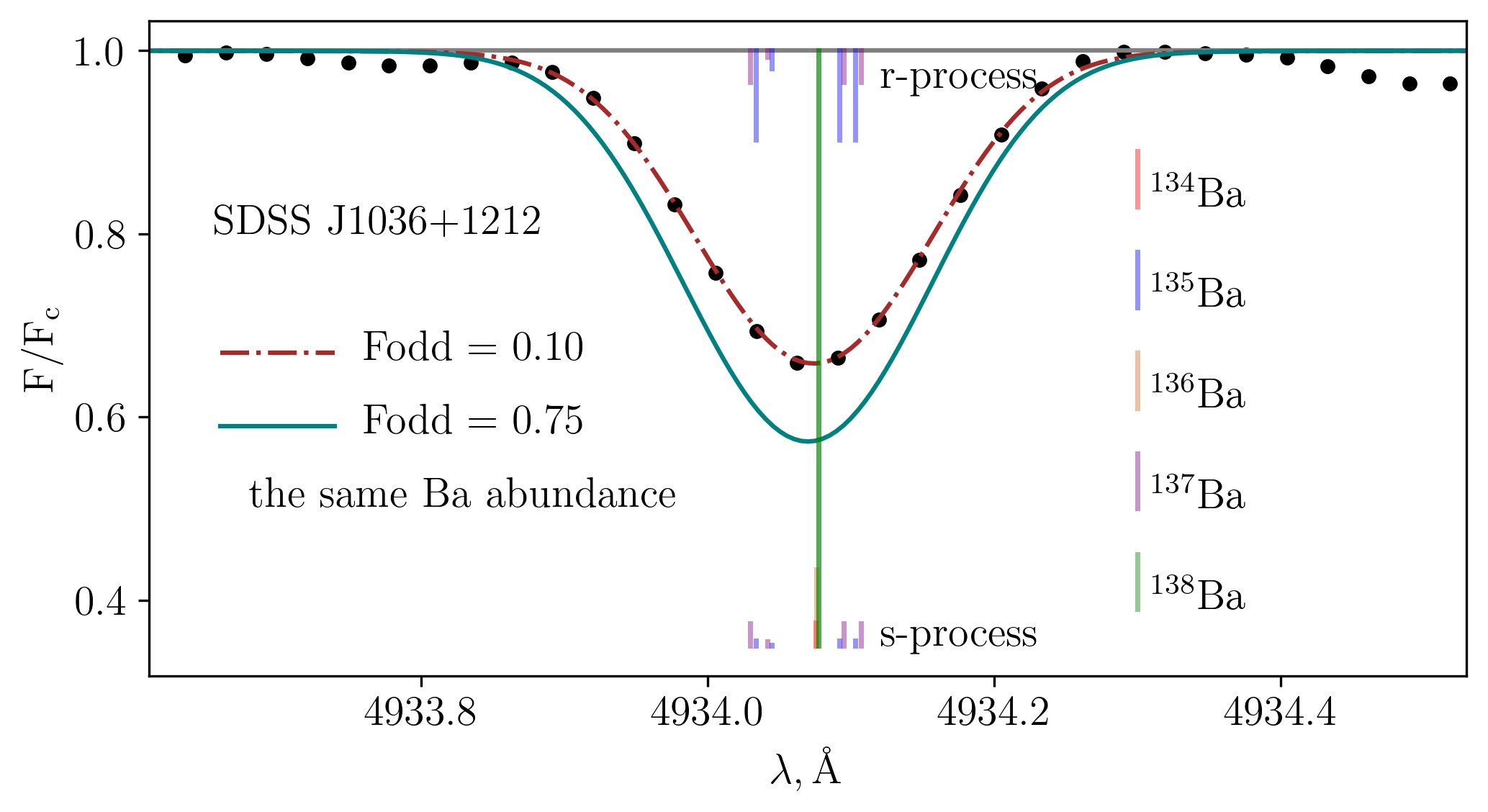}
\caption{The Ba\ii\ 4934 \AA\ line profiles in the observed spectrum of  \sjten\ (circles)  along with the synthetic NLTE spectra computed with the same Ba abundance but different \fodd\ = 0.10 and 0.75. Vertical dashes show the relative contribution of different Ba isotopes to the r-process and s-process; see legend for isotopes designations.
\label{ba_res_rs_same_abun}}
\end{figure}

\begin{figure}[H]
\centering
\includegraphics[trim={0 42 0 0},width=\hsize]{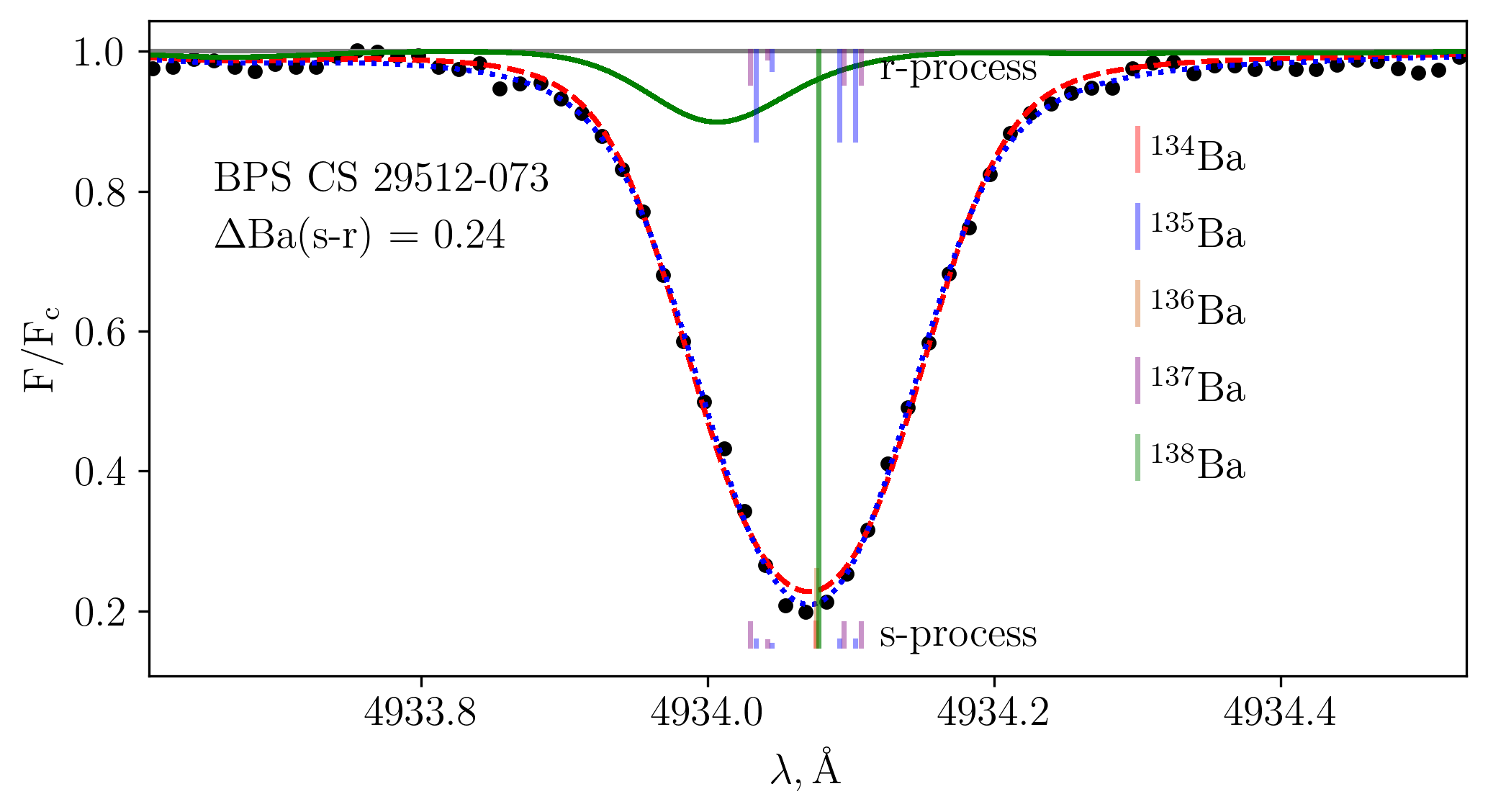}
\includegraphics[trim={0 42 0 0},width=\hsize]{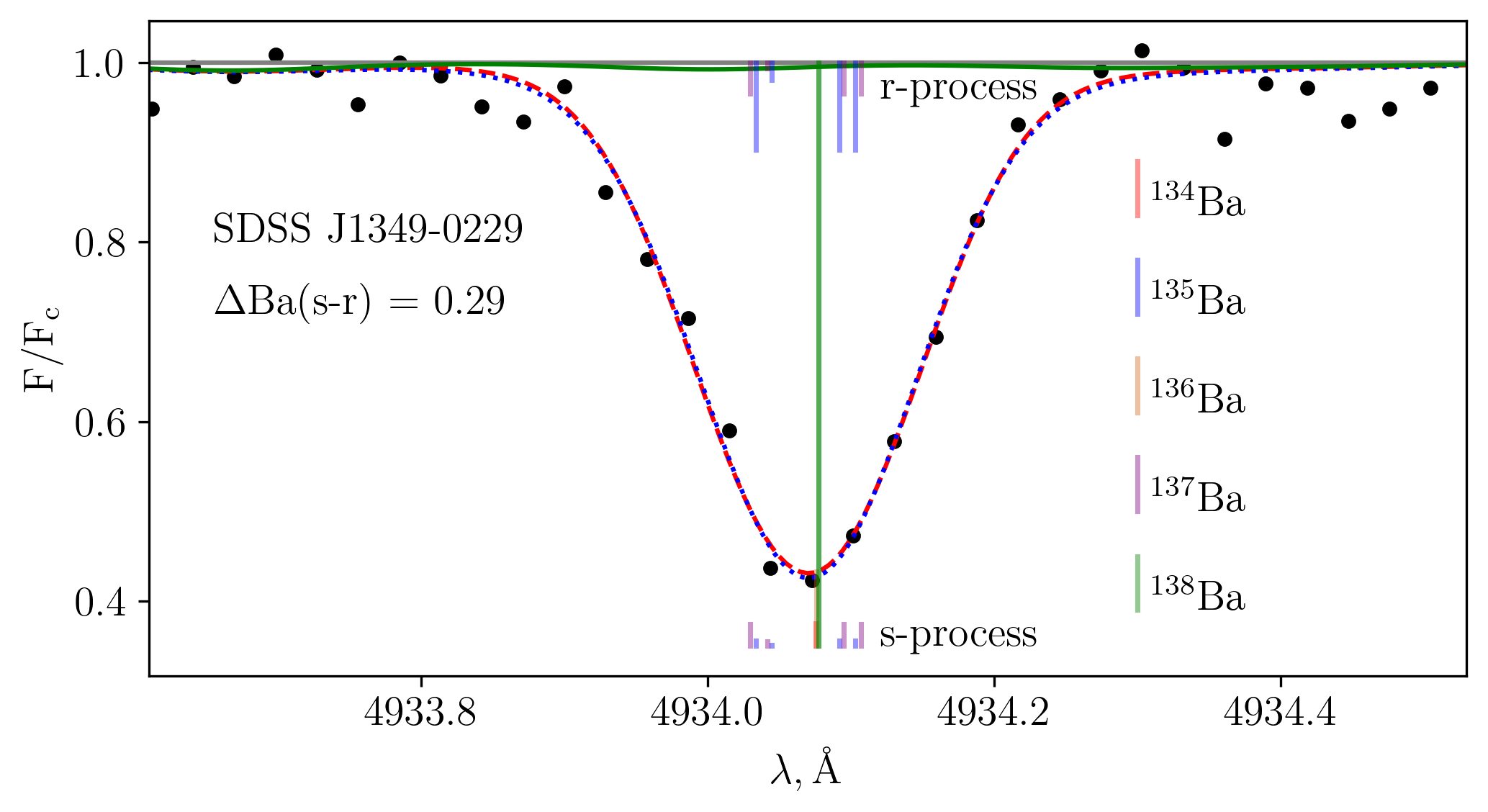}
\includegraphics[trim={0 10 0 0},width=\hsize]{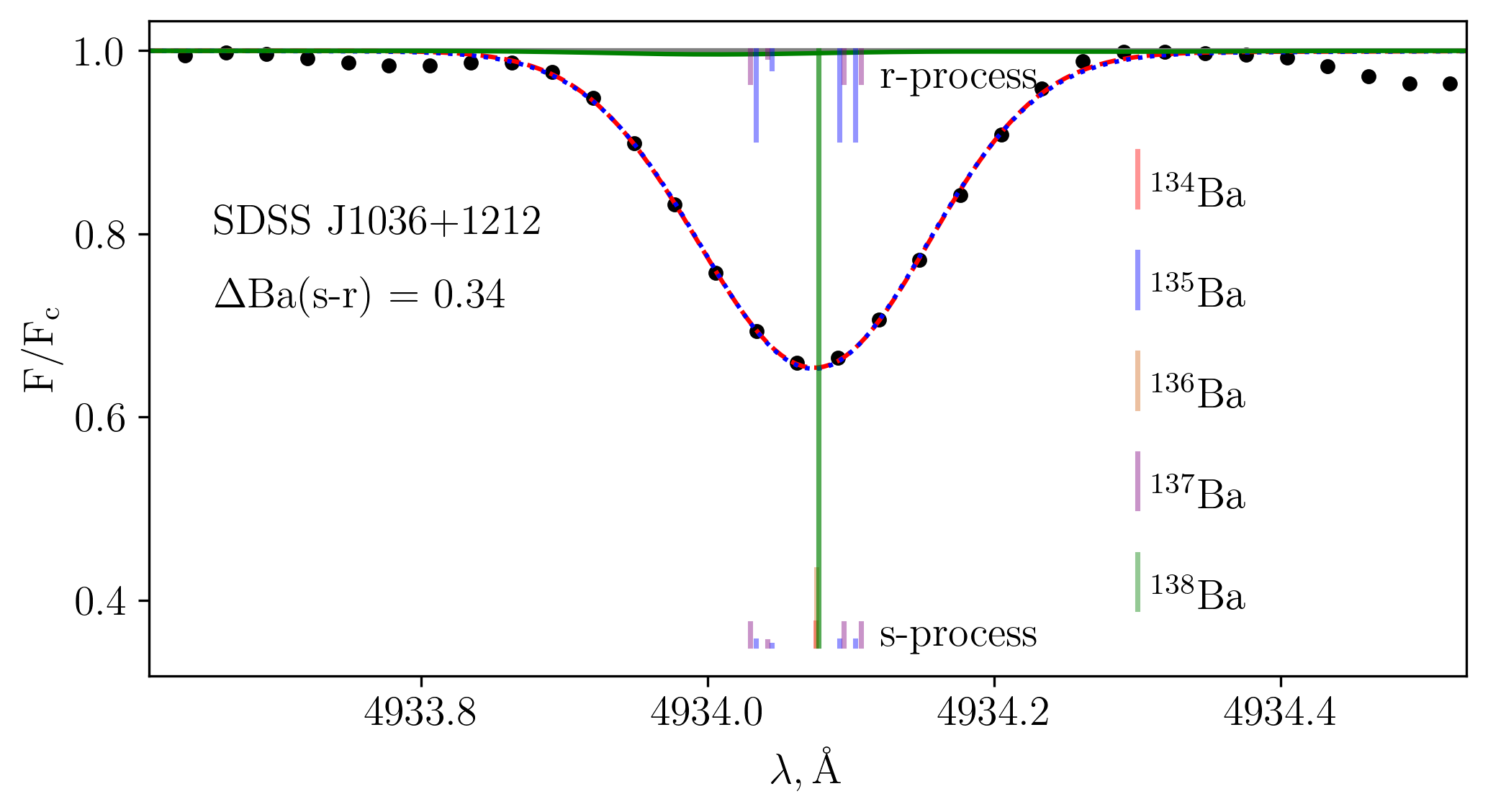}
\caption{Ba\ii\ 4934 \AA\ line profiles in the observed spectra (circles) of the sample stars along with the synthetic best-fit NLTE spectra, derived using pure r-process (dashed red line) and pure s-process (dotted blue line) Ba isotope mixtures. In the CEMP-s star BPS~CS~29512-073, the s-process Ba isotope mixture provides a better fit compared to those computed with the r-process mixture. In SDSS~J1349-0229 and SDSS~J1036+1212, the Ba\ii\ 4934~\AA\ lines are weaker compared to that in BPS~CS~29512-073 and their best-fit spectra derived with the s- and r-process Ba isotope mixtures are almost indistinguishable. For each star, the abundance difference between the best-fit spectra derived with the s- and r-process Ba isotope mixtures ($\Delta$Ba(s-r)) are indicated. Solid green line shows a spectrum, calculated neglecting the Ba\ii\ 4934 \AA\ line; the blending lines are attributed to the Fe\ione\ 4934.01~\AA\ and the CH 4933.70~\AA\ lines. Vertical dashes show the relative contribution of different Ba isotopes to the r-process and s-process; see legend for isotopes designations.
\label{fit_ba4934}}
\end{figure}

\begin{figure}[H]
\centering
\includegraphics[trim={0 10 0 0},width=\hsize]{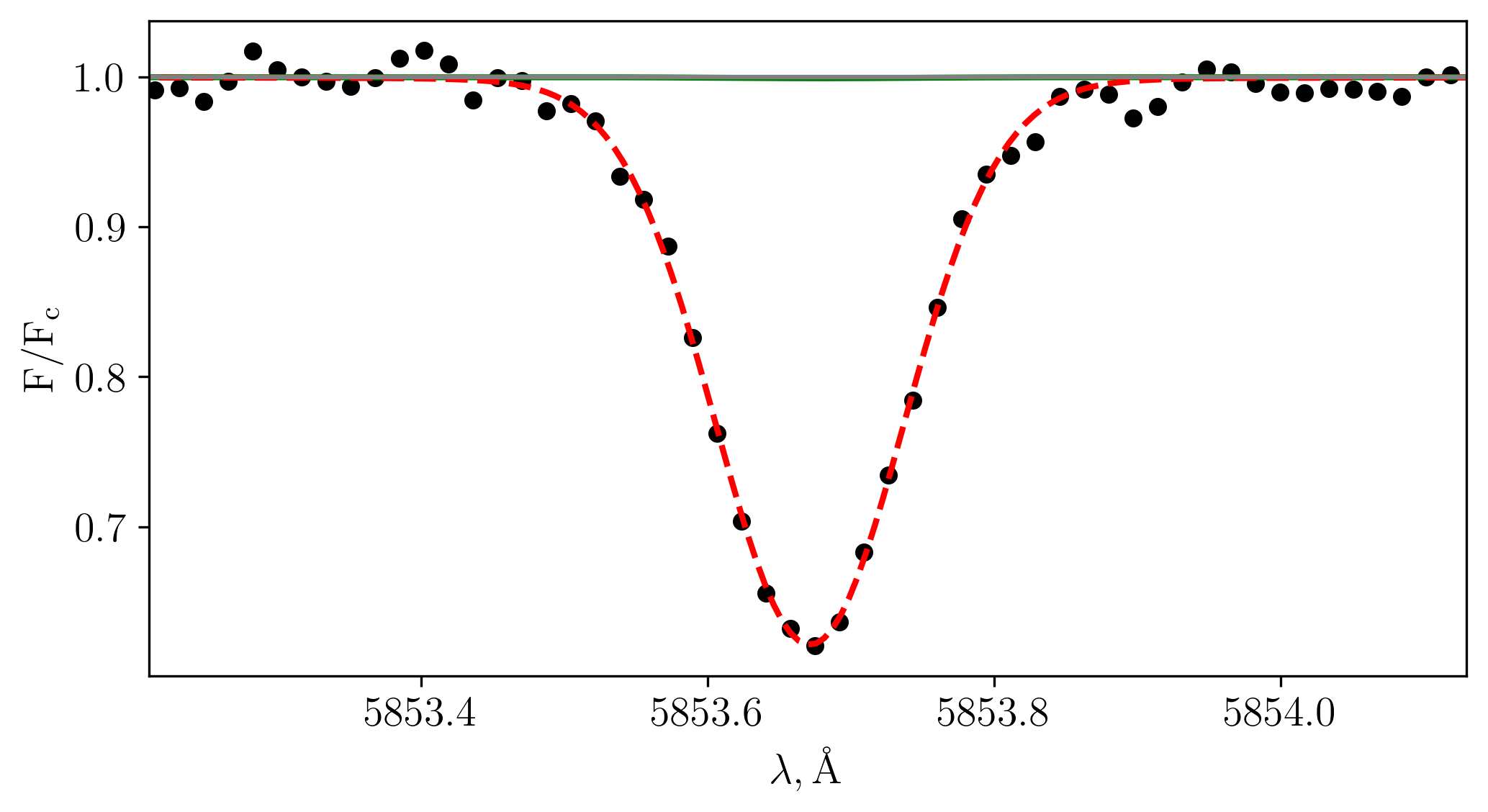}
\includegraphics[trim={0 10 0 0},width=\hsize]{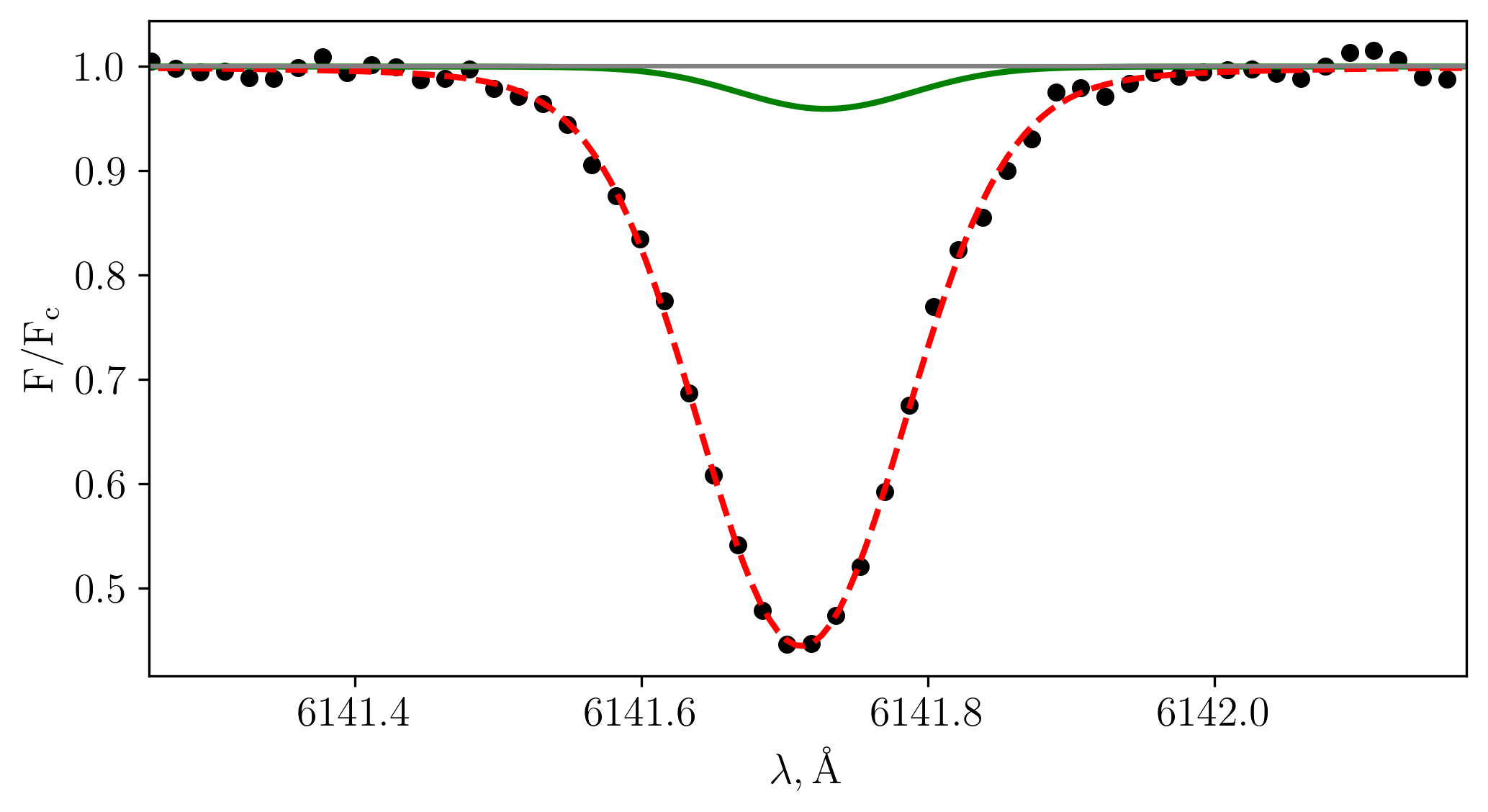}
\includegraphics[trim={0 10 0 0},width=\hsize]{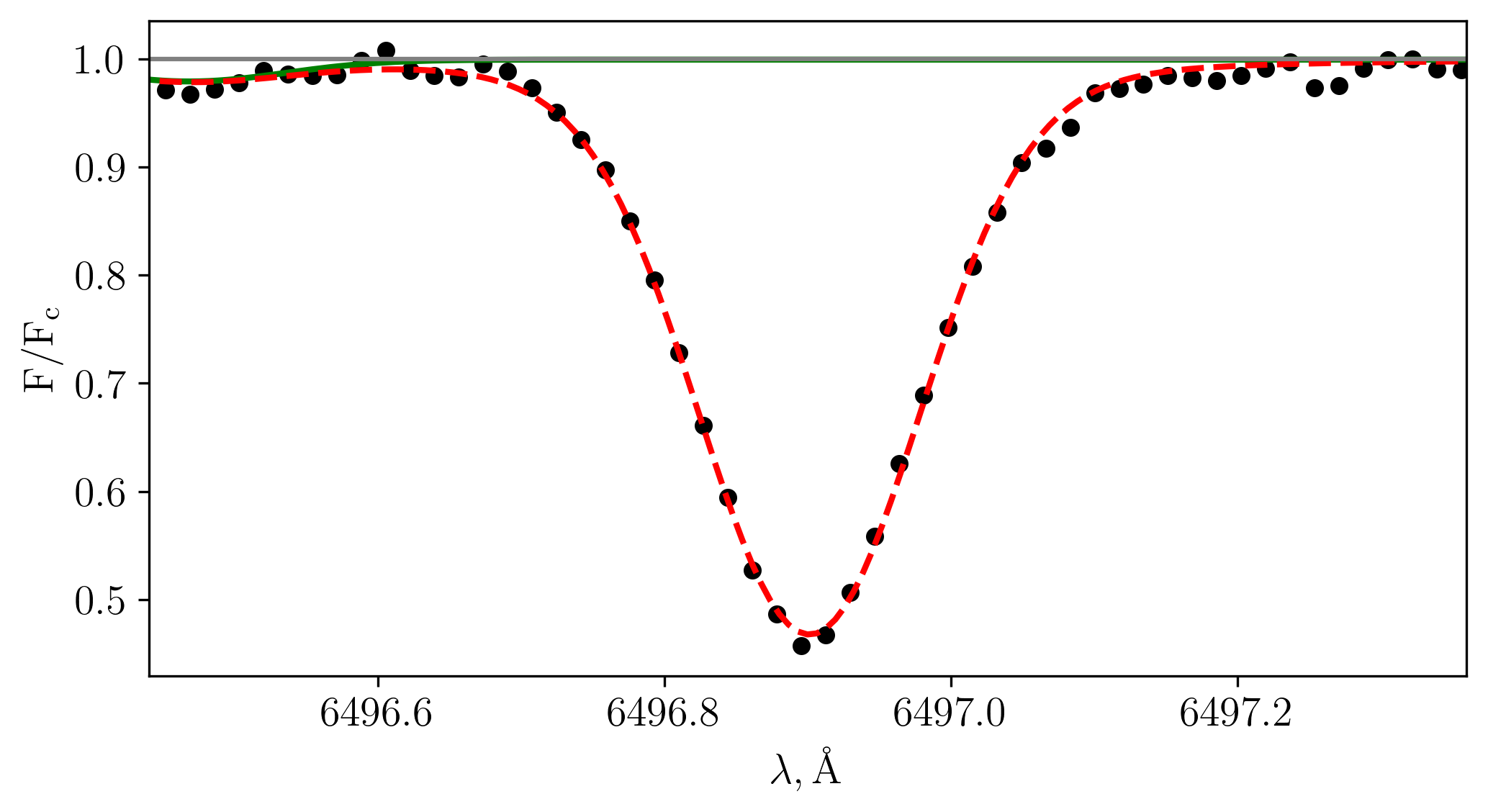}
\caption{Ba\ii\ line profiles in the observed spectra (circles) of the \cs\ stars along with the synthetic best-fit NLTE spectra (dashed line). Solid green line shows a spectrum, calculated neglecting the Ba\ii\ lines; the blending line is attributed to the Fe\ione\ 6141.73~\AA\ line (middle panel). 
\label{fit_ba_sub}}
\end{figure}

\begin{figure}[H]
\centering
\includegraphics[trim={0 10 0 0},width=\hsize]{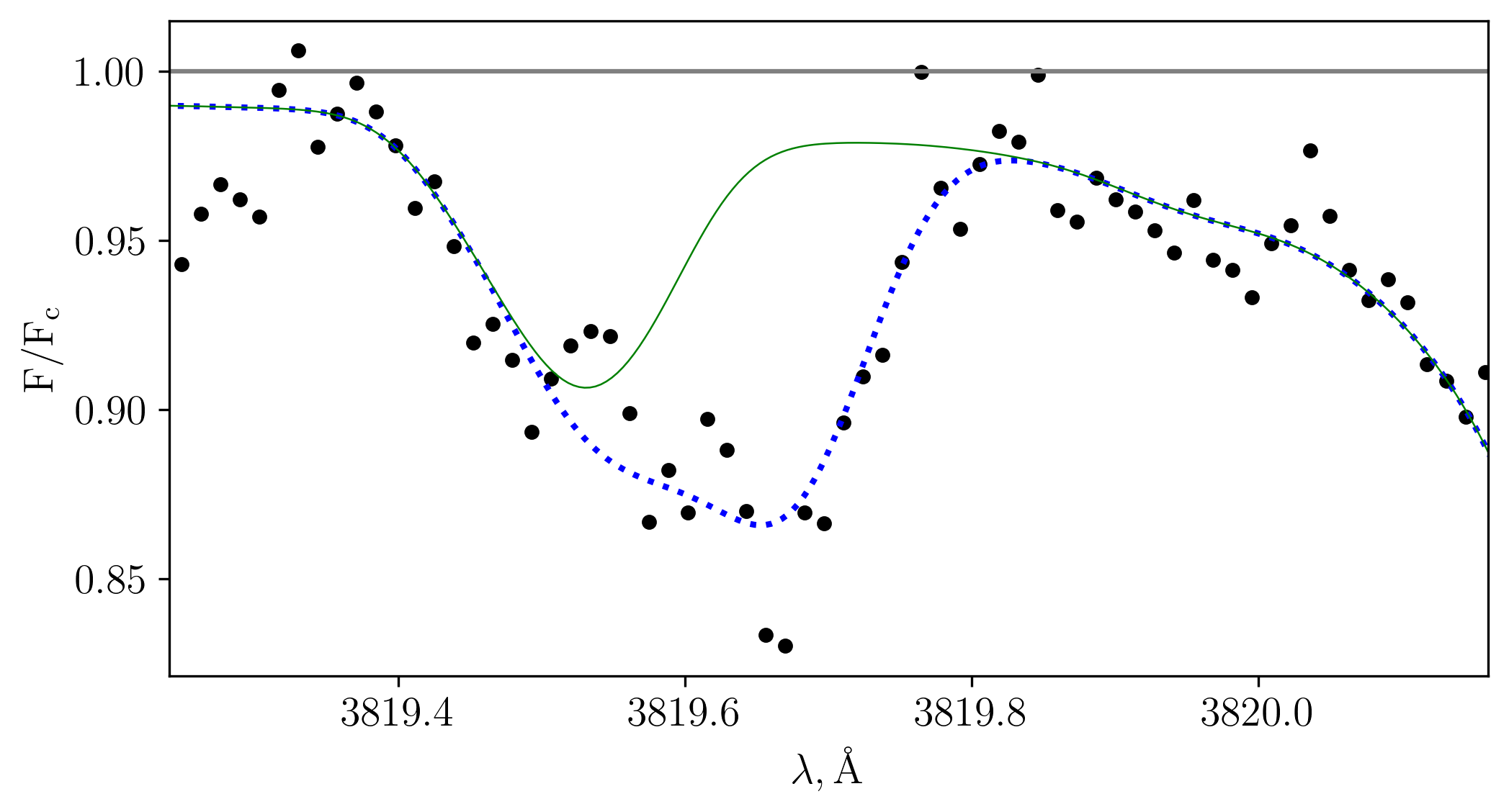}
\includegraphics[trim={0 10 0 0},width=\hsize]{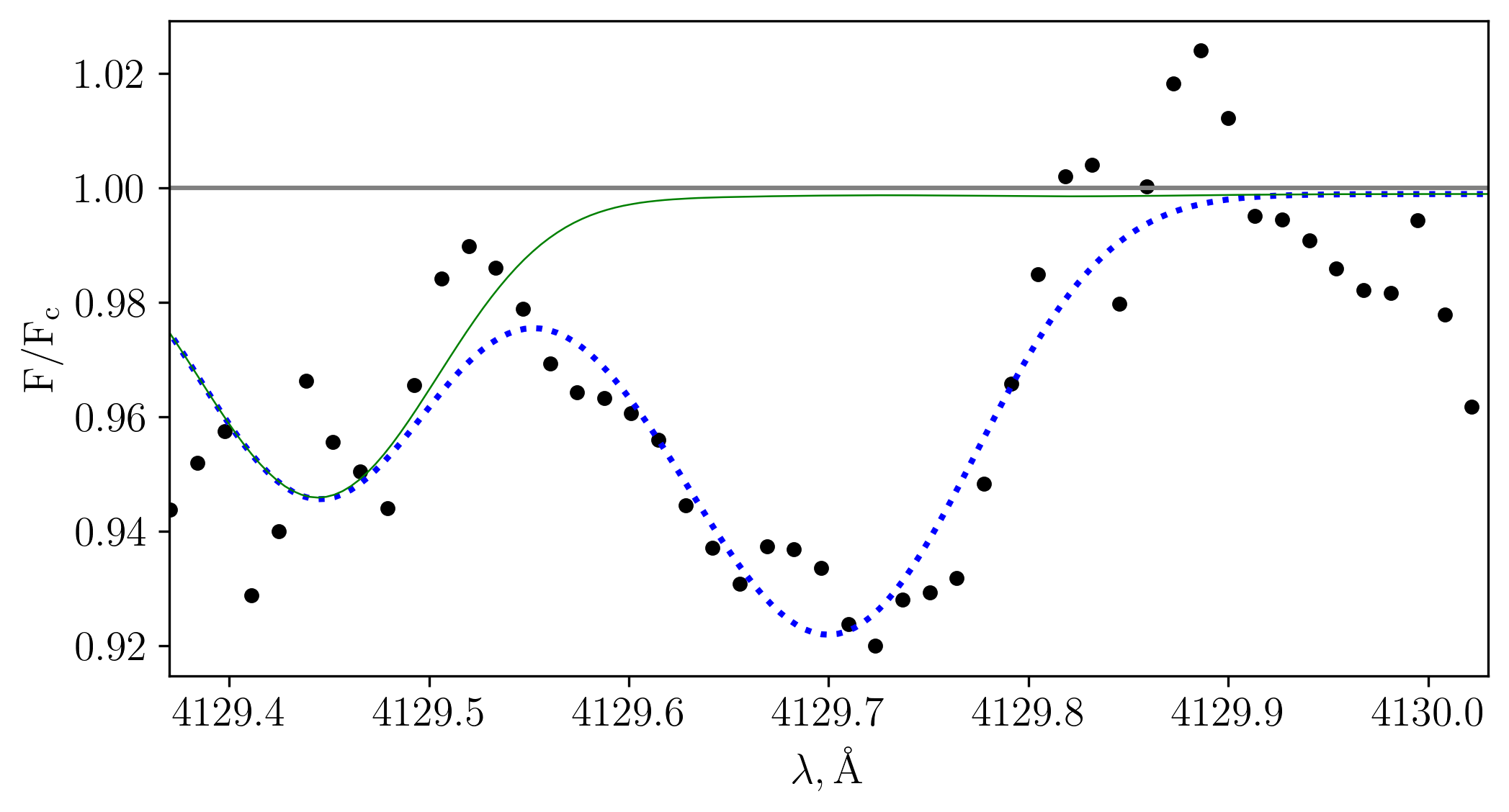}
\includegraphics[trim={0 10 0 0},width=\hsize]{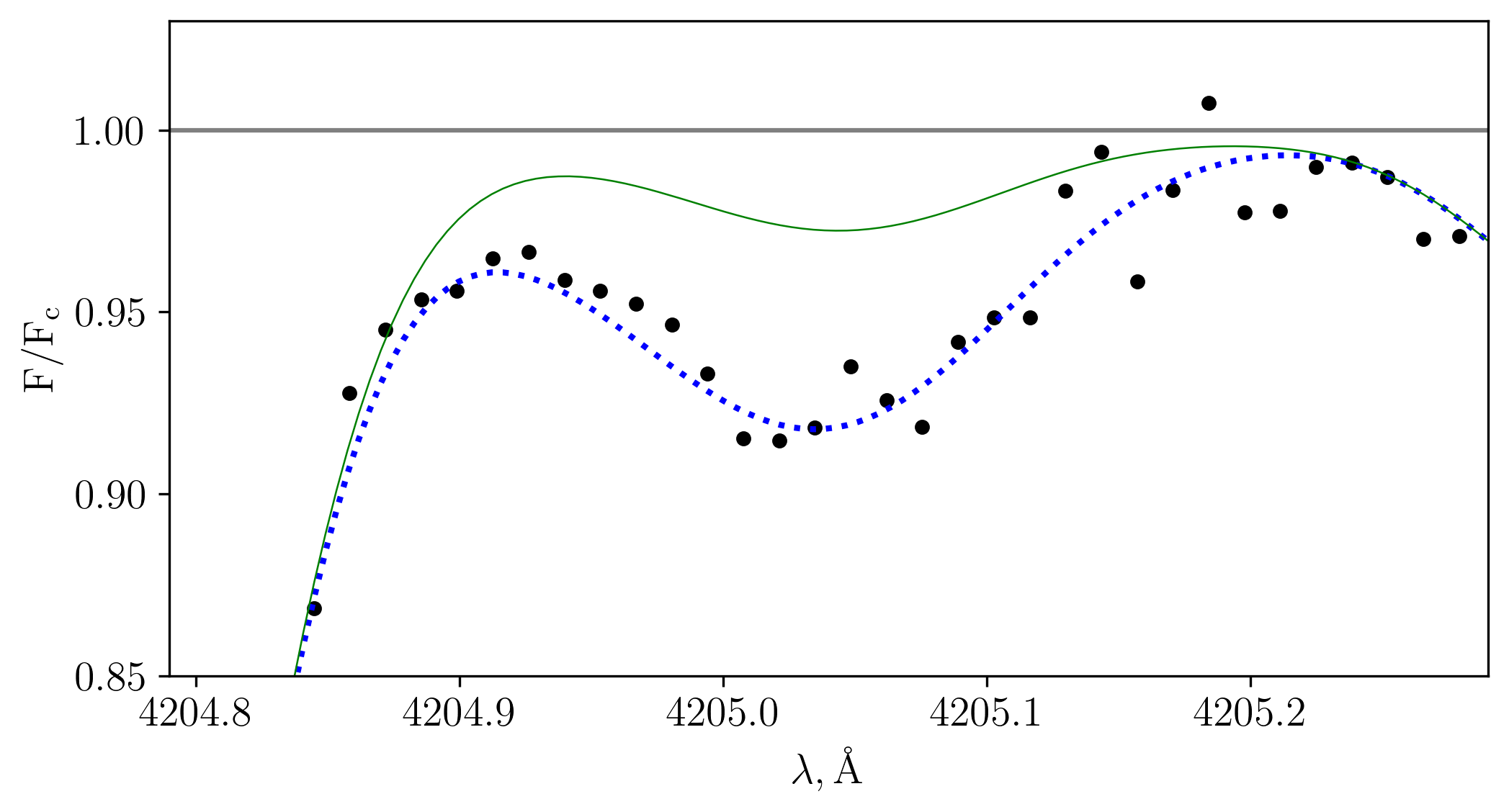}
\caption{Eu\ii\ line profiles in the observed spectra (circles) of the \cs\ stars along with the synthetic best-fit NLTE spectra (dotted line). Solid green line shows a spectrum, calculated neglecting the Eu\ii\ lines; the blending lines are attributed to the Fe\ione\ 3819.49~\AA\, Cr\ione\ 3819.56~\AA\ (top panel), Fe\ione\ 4129.26~\AA\ (middle panel),  CH 4204.70~\AA, and the V\ii\ 4205.05~\AA\ (bottom panel) lines. 
\label{fit_eu}}
\end{figure}


\begin{adjustwidth}{-\extralength}{0cm}




%


\PublishersNote{}
\end{adjustwidth}
\end{document}